\documentclass[11pt]{article}

\usepackage{times}
\usepackage{amsmath,amsfonts,amssymb,amsthm}
\usepackage{graphicx}
\usepackage{hyperref}    
\usepackage{color}
\usepackage{fancyhdr}
\usepackage{setspace}         
\usepackage{bm}  

\usepackage{threeparttable}
\newcommand{\beq}[1]{\begin{equation}\label{eq:#1}}
\newcommand{\eeq}{\end{equation}}
\newcommand{\beqar}[1]{\begin{eqnarray}\label{eq:#1}}
\newcommand{\eeqar}{\end{eqnarray}}
\newcommand{\refeq}[1]{Eq.\ ($\ref{eq:#1}$)}
\newcommand{\reffig}[1]{Fig.\ ($\ref{fig:#1}$)}
\newcommand{\av}[1]{\langle #1 \rangle}

\fancypagestyle{plain0}{
  \fancyhf{}
  \fancyhead[L]{{\footnotesize \sf A. Asztalos and Z. Toroczkai}}
  \fancyhead[C]{\footnotesize \sf }
  \fancyhead[R]{\footnotesize \sf Network Discovery by Generalized Random Walks}
 
  \fancyfoot[C]{\thepage}
}

\begin{document}

\title{Network Discovery by Generalized Random Walks }

\author{{\bf A. Asztalos}$^{1,2}$  and {\bf Z. Toroczkai}$^{1}$\thanks{E-mail: toro@nd.edu}\\
{$\mbox{\;\;\;}$}\\
{ $^{1}$ Interdisciplinary Center for 
Network Science and Applications (iCeNSA) and }\\
{ Department of Physics, University of Notre Dame, Notre Dame, IN, 46556 USA} \\
{ $^{2}$Department of Computer Science and  Department of Physics, }\\ 
{ Rensselaer Polytechnic Polytechnic  Institute, Troy, NY 12180-3590, USA}}

\maketitle

\begin{abstract}
We investigate network exploration by random walks defined via stationary
and adaptive transition probabilities on large graphs. We derive Ê
an exact formula valid for arbitrary graphs and arbitrary walks with 
stationary transition probabilities (STP), for the average number of discovered edges as 
function of time. We show that for STP walks  
site and edge exploration obey the same scaling $\sim n^{\lambda}$ as function of time $n$. 
Therefore, edge
exploration on graphs with many loops is always lagging compared to site exploration, the 
revealed graph being sparse until almost all nodes have been discovered. 
We then introduce the Edge Explorer Model, which presents a novel class of adaptive walks,
that perform
faithful network discovery even on dense networks.
\end{abstract}

\section{Introduction} 

Random walk theory \cite{RandWalksRandEnvVol1_H95, 
IntrProbTheorApplVol2_F71, RevMarkChRandWalksGraphs_AF99} has 
seen myriad applications ranging from physics,  biology and ecology 
through market models, finance, to problems in mathematics and computer 
science. It is being used
to sample distributions, compute volumes and solve convex optimization 
problems \cite{CombCompGeom_V05} and has played a key role
in www search engines  \cite{Proc8IntWWW_HHMN99}. 
It provides a microscopic description for real-world transport 
processes on networks, 
such as information spread \cite{Holme_traffic03, PRL_Rieger04} and disease propagation 
(epidemics \cite{PRL_epidemic01, Phys_Rev_E_N02, PRE_Volchenkov02}) and  
it can also be used to design  network discovery/exploration tools \cite{PRE_Adamic01}.  
Here  we focus on the latter aspect. 

The structure of real-world networks \cite{Strogatz_Nature01,StatMech_BA02} is 
organically evolving and frequently, either their size 
is simply prohibitive for measuring their topology (the WWW has
$\sim 2\times 10^{10}$ nodes), or the nature 
of the network makes it difficult to gather global information (e.g., in some social networks). 
Such networks are best explored by `walkers' stepping from a node to a neighbor
node linked by an edge, collecting local information, which is 
then combined to produce subgraph samples of the original graph. 
To fix notations, we denote by $G(V,E)$ the graph on which the walk happens,  
where $V$ is the set of $N$ nodes, $E$ is the set of $M$ edges, and by $p(s'|s;t)$ 
the single-step transition probability of a walker at site (node) $s$ to step onto 
a neighboring site $s'$ ($(s,s') \in E$), on the $t$-th step. 
Note that equivalently, one could consider the walk taking place on the complete 
graph $K_N$, with setting $p(s'|s;t) = 0$ for $(s',s) \not\in E$.
One can think of $p(s'|s;t)$ as information `handed' to the walker at node $s$ to 
follow in its choice for stepping to the next node.  Accordingly, an important
optimization problem is  to `design' the $p(s'|s;t)$ 
probabilities such that certain properties of the exploration are optimal. 
Such problems motivate the development of the statistical 
mechanics of network discovery, connecting the set of local transition probabilities 
$\{p(s'|s;t)\}$ with the global properties of the uncovered subgraph as function of time.
We distinguish two main classes of exploration problems, namely those with: 
I. stationary transition probabilities (STP) where $p(s'|s;t) = p(s'|s)$ 
(time-independent) and II. adaptive transition probabilities (ATP) where $p(s'|s;t)$ 
depends on time and possibly on the past history of the walk. 
For general STP walks, analytic results were obtained for
the number $S_n$ of distinct (virgin) nodes visited in $n$ steps (site exploration) 
 \cite{DvoEr51,MW65,Cassi_review05} (for a review see \cite{RandWalksRandEnvVol1_H95}),  
 and on the cover time $T_V^*$ (expected number  of steps 
to visit all nodes, for a review see \cite{Lovasz_93Survey}).
Numerically, site exploration by simple random walks $p(s'|s) = k_s^{-1}$ ($k_s$ is the
degree of node $s$) has been extensively studied on various complex  network models 
\cite{PRE_Lahtinen01, PRE_Eivind03}. 

\begin{figure}[htbp]
\centerline{\includegraphics[width=3.7in]{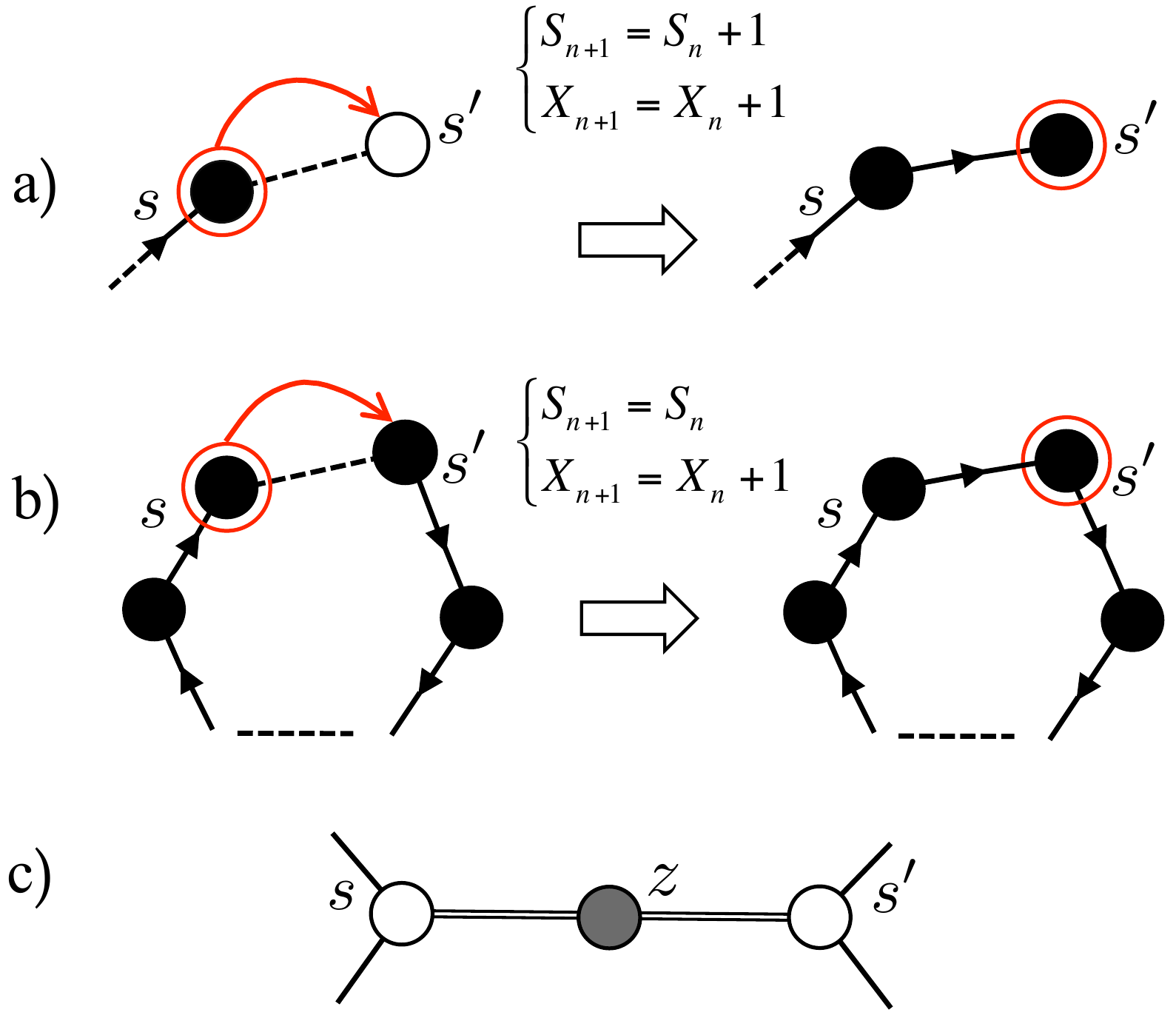}} 
\caption{a) A new node (empty circle) is always discovered via a new edge (dashed line). 
b) A new edge,  can also be discovered between already visited nodes. c) First passage
time through an edge can be computed from the first passage time to a pseudonode $z$
placed on the edge.} 
\label{fig:site_node_expl_pseudonode}
\end{figure}

Interestingly, the number $X_n$ of distinct  {\em edges} (edge exploration) visited in 
$n$-steps has only been studied numerically, for simple random walks, 
\cite{EdgeEx_CS_Feige93, EPL_Ram07}, and no analytic results similar to $S_n$ 
have been derived. The statistics of $X_n$, however, cannot  be obtained directly 
from the analytic results for $S_n$ on a ``dual'' graph, such as the edge-to-vertex 
dual graph $L(G)$, because the walk does not transform simply onto $L(G)$. 
On graphs with loops, there
is an inherent asymmetry between the evolution of $S_n$ and $X_n$. While a new
node is always discovered via a new edge ($S_{n+1} = S_n +1$ implies 
$X_{n+1} = X_{n} + 1$, \reffig{site_node_expl_pseudonode}a)), a new edge can
be discovered between two previously visited nodes as well
(\reffig{site_node_expl_pseudonode}b)). In the latter case, the 
walker always encloses a loop in the discovered subgraph, hence the
$(S_n,X_n)$ pair can be connected to the {\em loop statistics} of the network. More
precisely, the quantity $Q_n = 1+X_n-S_n$ gives the number of times the walker returned
to its own path through a freshly discovered edge, in $n$ steps. 
Clearly, if $G$ is a tree, then $X_n = S_n-1$ for all $n \geq 0$. 

In this Letter we provide an exact expression for the generating function for 
the average number of discovered edges $\langle X_n \rangle$ in $n$-steps 
for arbitrary STP walks on arbitrary graphs. Although our expressions are valid
in general, we are interested in the {\em scaling behavior} of $\langle S_n \rangle$
and $\langle X_n\rangle$ for large times ($n \gg 1$) on large  ($N \gg 1$) 
{\em connected} graphs. Let us consider a monotonically increasing sequence
$\{c_n\}$ of positive terms $c_n > 0$. We will use the notation 
$c_n \sim n^{\kappa}$ to say that $c_n$ scales with $n$ with a growth 
exponent $\kappa$, if as $n \to \infty$, $c_n \simeq n^{\kappa} L(n)$, where
$L(n)$ is a slowly varying function that is
$L(\eta x)/L(x) \to 1$ as $x \to \infty$ for any $\eta > 0$.

As it will be seen, for STP walks on large but finite graphs both 
the average number of discovered nodes and edges obey scaling laws
$\langle S_n \rangle \sim n^{\lambda}$, $ \langle X_n \rangle \sim n^{\mu}$. 
These hold up to a {\em cross-over time} $T_V$ (for $\langle S_n \rangle$)
and $T_E$ (for $\langle X_n \rangle$) after which {\em saturation} sets in until all
the nodes (edges) have been discovered, at the corresponding
cover times $T_V^*$ and $T_E^*$. At the cross-over time only a small {\em constant 
number} of nodes (edges) are left untouched and we consider this as the time
where the discovery has practically been completed. 
Since in a step at most one node (edge) can be discovered, 
the growth of $S_n$ ($X_n$) is {\em at most linear},
at any time, for any walk (STP or ATP), that is $\lambda \leq 1$, $\mu \leq 1$.  
When the walker is in completely charted territory, both $S_n$ and $X_n$ stagnate, 
otherwise $X_n$ always grows (\reffig{site_node_expl_pseudonode}a-b) ), and hence
$\mu \geq \lambda$. 

Here we show that for {\em recurrent} STP walks both site and edge 
exploration obey the same scaling that is $\lambda =\mu$ in the 
$N \to \infty$  limit. 
This means that  in dense graphs where the nr of edges $M \sim N^{\nu}$, 
$\nu > 1$  (but $\nu \leq 2$),  the node set $V$ is discovered much earlier than 
the edge set $E$ ($T_V < T_E$). As we prove below, even for the complete 
graph on $N$ nodes 
$G = K_N$, STP walks explore nodes and edges at the same rate, $\mu = \lambda$. 
This is counterintuitive, because there are 
${\cal O}(N^2)$ edges, and the walker could keep discovering many new edges between
already visited nodes, so there is no obvious reason why we could not have $\mu > \lambda$.
The fact that for STP walks, edge and site exploration grow at the same rate, presents
a problem if one is interested in discovering the {\em links} (relationships) in a network. 
Moreover, if network
discovery is done with the purpose of sampling and producing a subgraph with statistical
properties resembling that of the underlying network, then STP walks will not provide
the optimal solution, independently on the form of the transfer matrix $p(s|s')$.  This is 
simply because a walker's choice to move to a neighbor will be independent of its visiting 
history, and therefore will have a lower chance on average to discover a virgin edge 
to a visited neighbor (Fig. \ref{fig:site_node_expl_pseudonode}b))  than for e.g., an 
ATP walk that is biased towards already 
visited neighbors. Hence,  for a given number of visited nodes in an STP walk 
$\langle X_n \rangle = {\cal O} (\langle S_n \rangle)$ number of edges will be revealed 
before $T_V$, making
the discovered subgraph sparse, seriously skewing the sample especially, if the underlying
network is dense ($\nu > 1$).  To resolve this, we 
introduce an ATP walk, the Edge Explorer Model (EEM) that performs a faithful
exploration of the  nodes and edges even on dense networks, such that $T_E \simeq T_V$. 

\section{STP walks}  The generating function  
$S(s_0;\xi) = \sum_{n=0}^{\infty}\langle S_n \rangle \xi^n$ for the average number of 
distinct sites discovered in $n$ steps by the walker starting from site $s_0$ 
can be written as \cite{RandWalksRandEnvVol1_H95}:
\begin{eqnarray}
S(s_0;\xi) &=&   
\frac{1}{1-\xi} \sum_{s \in V} W(s;\xi) P(s|s_0;\xi), \label{sx}\\
W(s;\xi) &= & \left[P(s|s;\xi)\right]^{-1}\;, \label{sxw}
\end{eqnarray}
where $P(s|s_0;\xi)$ is the site occupation
probability generating function, that is $P(s|s_0;\xi) = \sum_{n=0}^{\infty} \xi^n P_n(s|s_0)$,
with $p_n(s|s_0)$ being the probability for a walker starting from $s_0$ to be found 
at $s$ on the $n$-th step. Next we derive a similar expression for 
$X(s_0;\xi)=\sum_{n=0}^{\infty}\langle X_n \rangle\xi^n$.  
Let $F_n(s|s_0)$ be the first-passage time distribution (the probability for the walker
to arrive at $s$ {\em for the first time} on the $n$-th step) and let $F(s|s_0;\xi)$ be 
its generating function. It is well known \cite{RandWalksRandEnvVol1_H95} that 
$F(s|s_0;\xi)=\left[P(s|s_0;\xi) - \delta_{s,s_0}\right]/P(s|s;\xi)$. The probability that
$s$ is ever reached by the walker starting from $s_0$ is therefore $R(s|s_0) = 
\sum_{n=1}^{\infty} F_n(s|s_0) =\lim_{\xi \to 1, |\xi| < 1} F(s|s_0;\xi)$.  Since 
$R(s|s_0) \leq 1$, $F(s|s_0;1^{-})$ is convergent, however, $P(s|s_0;1^{-})$ can diverge.
If $R(s|s_0) = 1$ for all $s$, $s_0$, then the walk is {\em recurrent}, $P(s|s_0;1^{-}) = \infty$
and $F(s|s_0;1^{-})=1$. Moreover, in this case 
$P(s|s_0;1^{-})$ has the {\em same rate of divergence} for all $s,s_0 \in V$ 
\cite{RandWalksRandEnvVol1_H95}. For finite networks,
in which the walker can access all nodes and there are no traps, the walk is recurrent.

Let $F_n(e|s_0)$ denote the {\em edge} first-passage time distribution, i.e., the 
probability that the walker passes through edge 
$e=(s,s')\in E$ {\em for the first time} on the $n$-th step, given that it started at node $s_0$. 
 By introducing an indicator $\Gamma_n(s_0)$ for the number of virgin 
 edges discovered on the $n$th step ($=0,1$), we have 
$\av{\Gamma_n} =\mbox{Prob}\{\Gamma_n =1\}=\sum_{e\in E} F_n(e|s_0)$, 
with $\Gamma_0 = \langle \Gamma_0 \rangle=0$. Clearly, 
$X_n(s_0)=\sum_{j=0}^{n} \Gamma_j (s_0)$ and thus the generating function for 
the average number of visited distinct edges in $n$ steps becomes:
\begin{equation}
X(s_0;\xi) =\frac{\Gamma(s_0;\xi)}{1-\xi} = 
\frac{1}{1-\xi} \sum_{e \in E} F(e | s_0; \xi)\;, \label{Xxi_gamma}
\end{equation}
where $\Gamma(s_0; \xi)$ and $F(e|s_0; \xi)$ are generating functions 
for $\Gamma_n(s_0)$ and $F_n(e|s_0)$, respectively.

To obtain the edge first-passage time distribution, we  place an auxiliary site 
$z$ on edge $e=(s,s')\in E$ (Fig \ref{fig:site_node_expl_pseudonode}c)) and 
redefine the walk on this
new graph $G_z$ such that the {\em node} first-passage time probability to  $z$ 
on $G_z$ is the same as the {\em edge} first-passage time probability through $e$ on $G$. 
The extended graph $G_z(V_z,E_z)$ has  $V_z =V \cup \{z\}$ and 
$E_z=\{ (s,z); (z,s')\} \cup E \setminus \{ (s,s')\}$. The addition of 
$z$ to $e=(s,s')$ changes 
only the transition probabilities around that edge, leaving $p(r|r')$ 
the same away from $s$ and $s'$.
 Steps from $s$ ($s'$) to $s'$ ($s$) in the new walk are forbidden; 
 instead, the walker has to step onto node $z$ first. However, 
 the same probability flow has 
 to exist in the modified walk as in the original one when moving from sites $s$ and $s'$ 
 towards $z$. From $z$ the walker is only allowed to step to $s$ or $s'$ with arbitrary 
 probabilities $f$ and $g$, respectively (which, however, should not enter the final 
 expression for  $F(e | s_0; \xi)$!). The single-step transition probabilities on 
 the  $G_z$ for the new walk can thus be combined into ($r,r' \in V_z$):
 \begin{eqnarray}
p^{\dagger}(r|r')=(1-\delta_{r'z}-\delta_{rz}-\delta_{r's'}\delta_{rs} - 
\delta_{r's}\delta_{rs'})p(r|r')+&&  \nonumber \\
\delta_{rz}\delta_{r's'}p(s|s')+\delta_{rz}\delta_{r's}p(s'|s) +
+ \delta_{r'z}(f \delta_{rs} + g \delta_{rs'})\;.\;&& \label{eq:new_walk}
\end{eqnarray}
The rest of the calculation focuses on obtaining the node first-passage time distributions 
$F^{\dagger}_n (z|s_1)$ and site occupation probabilities 
$P^{\dagger}_n (r|s_1)$ of the modified walk
($r,s_1 \in V_z$). Due to our setup we have
$F_n(e|s_0) = F^{\dagger}_n(z|s_0)$, $s_0\in V$, or 
$F(e|s_0;\xi) =F^{\dagger} (z|s_0;\xi) =P^{\dagger} (z|s_0;\xi)/P^{\dagger} (z|z;\xi)$,
$s_0 \neq z$. 
Obtaining the $P^{\dagger}$ generating functions in terms of the original functions
$P$ involves a lengthy series of Green-function manipulations, using the formalism developed for 
`taboo sites' \cite{RandWalksRandEnvVol1_H95,Chung1960}, see Appendix A. 
The final result after using (\ref{Xxi_gamma}) is:
\begin{equation}
X(s_0;\xi)=\frac{\xi}{1-\xi} \sum_{s\in V} \overline{W}(s;\xi) P(s|s_{0};\xi)\;, \label{X_gf}
\end{equation}
with
\begin{equation}
\overline{W}(s;\xi)=\sum_{s' \in V}\alpha \frac{1+(d-c)\beta \xi}{1+(a \alpha +d \beta)\xi + 
(ad-bc) \alpha \beta \xi^{2}}\;, \label{W}
\end{equation}
where $a=a(\xi) = P(s|s';\xi)$, $b = b(\xi) = P(s|s;\xi)$, $c = c(\xi) =P(s'|s';\xi)$, 
$d =d(\xi) =P(s'|s;\xi)$ and $\alpha=p(s'|s)$, $\beta=p(s|s')$. 
The form (\ref{X_gf}) is similar to (\ref{sx}), however with a more involved weight
function $\xi\overline{W}(s;\xi)$. This shows that
edge exploration is usually quite different from node exploration.  
Expressions (\ref{X_gf}-\ref{W}) are entirely general, valid
for any type of STP walk (including asymmetric walks $p(s|s') \neq p(s'|s)$ ), on arbitrary 
graphs. The properties
of $\xi b(\xi)\overline{W}(s;\xi)$ fully determine the statistics of edge exploration,  
{\em when compared} to site exploration. Note that the summation in the expression
of the weight $\overline{W}(s;\xi)$ is only over the network neighbors of $s$, due to the 
multiplicative transition probability $\alpha = p(s'|s)$, which is zero if $s'$ and $s$ are 
not neighbors in the graph. Next, we discuss some special cases for  simple random walks.  

\section{Special cases} 
\begin{figure}[htbp]
\centerline{\includegraphics[width=5.5in]{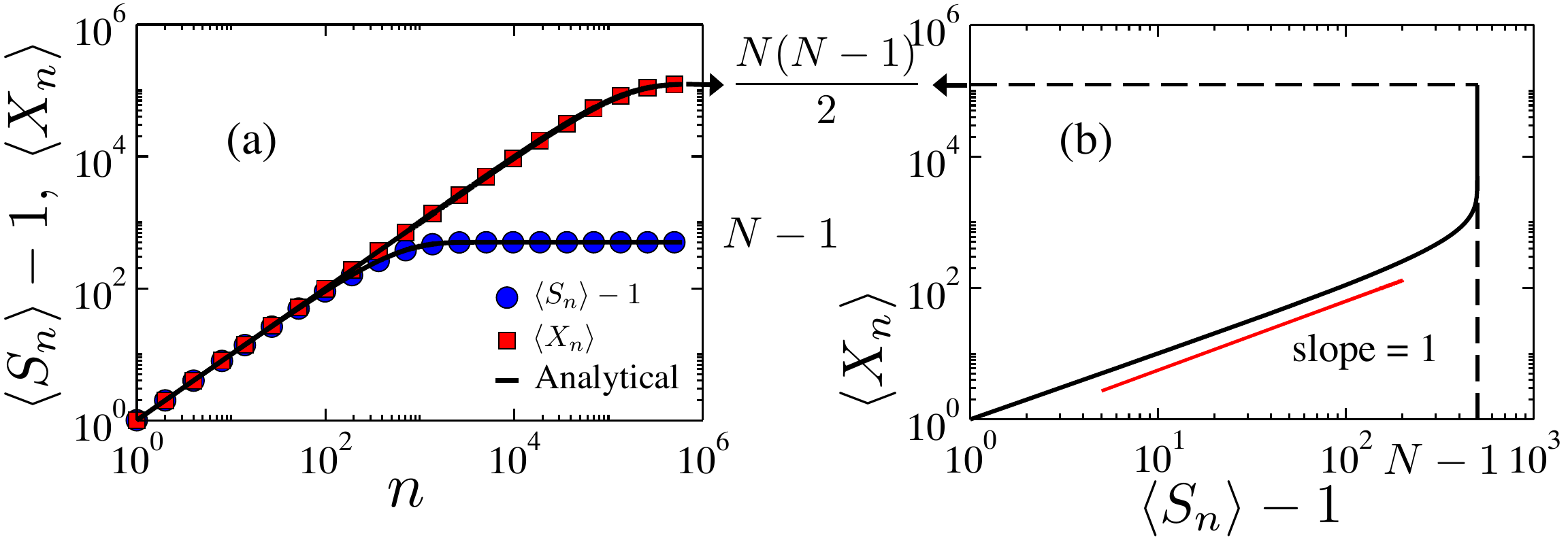}} 
\caption{(a) Numerical (blue circles for $\av{S_n}$, red squares for 
$\av{X_n}$, respectively) and analytical (black solid line) results shown 
for a complete graph of $500$ nodes. (b) Same as a) with $\av{X_n}$ 
vs $\av{S_n}$, showing that the discovered graph is sparse
even on $K_N$. $\av{S_n}-1$ is plotted instead of $\av{S_n}$ 
in order to account for the initial conditions $S_0=1$, $X_0=0$.} 
\label{fig:CG_numerics}
\end{figure}

{\em I. Simple random walks on $K_N$.} The
single-step probabilities of a simple random walker can be written as 
$p(s|s')=p(1-\delta_{ss'})$, $p=1/(N-1)$.  In this case 
$$P(s|s_0; \xi) = \left[\delta_{s,s_0} + p\xi (1-\xi)^{-1} \right](1+p \xi)^{-1}\;,$$ and $S(\xi)$ and 
$X(\xi)$ are easily obtained. In particular, $$S(\xi) = (1-\xi)^{-1} (1+p \xi) 
\left[ 1- (1-p)\xi\right]^{-1}\;,$$ from where, via contour integration, $\langle S_n \rangle  = 
\left[ 1+p-(1-p)^n \right]/p$ . 
Similarly, $$\overline{W}(s;\xi)= \left[ 1+(a+b)p\xi \right]^{-1},$$ and since in this case $\xi a = b -1$, 
we have
$$\overline{W}(s;\xi)= \left[ 1-p + (1+\xi)p b \right]^{-1}.$$ 
From (\ref{X_gf}) and the
expression for $b=P(s|s;\xi)$ it follows:
\begin{equation}
X(\xi) = \frac{\xi}{1-\xi} \cdot \frac{1-\xi +Np\xi}{1+(2p-1)\xi + 2p(p-1)\xi^2}\;. \label{xikn}
\end{equation}
After contour integration we obtain the exact expression
\begin{equation}
\av{X_n} =  \frac{1+p}{2p^2} +\frac{q_{1}q_{2}}{q_{1}-q_{2}} \left[ 
\frac{1+p q_{1}}{q_{1}^{n}(q_{1}-1)} - \frac{1+p q_{2}}{q_{2}^{n}(q_{2}-1)}\right], 
\label{Xn_cgg} 
\end{equation}
$ n \geq 0$.  Here $q_{1} > 1$, $q_2 < -1$ are the roots 
of the quadratic equation $2p(1-p)\xi^2 - (1-2p)\xi -1 = 0$.
\reffig{CG_numerics}a) shows the agreement between simulations and the analytical 
formulas for $\av{S_n}$, $\av{X_n}$.  From the above, for large graphs, $p\ll 1$,
$\av{S_n} - 1=p^{-1}\left[ 1-(1-p)^n \right]=n - {n \choose 2}p+\ldots$, showing
that $\av{S_n} \sim n$ in the regime $np \ll 1$, or $n \ll N$. Similarly, for large graphs,
$q_1 = 1+ 2p^2 + {\cal O}(p^3)$, $q_2 = -\frac{1}{2p}-\frac{1}{2} + {\cal O}(p)$, yielding
$\av{X_n} \sim (1+p) (n-1 + \ldots)$ for  $n \ll N^2$. 
The cover times can also be calculated, yielding $T^*_V \sim N \ln{N}$
(coinciding with \cite{Lovasz_93Survey})  and 
$T^*_E \sim N^2 \ln{N}$. Thus, both $\av{S_n}$ and $\av{X_n}$ grow {\em together}, linearly, 
($\lambda = \mu = 1$) as function of time, with $\av{S_n}$ saturating to $N$ at $T_V$ 
and $\av{X_n}$ to $N(N-1)/2$ much later, at $T_E$.  \reffig{CG_numerics}b) shows the linear dependence of $\av{X_n}$ on $\av{S_n}$.

{\em II. Infinite translationally invariant lattices in $d$ dimensions.} Simple 
random walks on such graphs are homogeneous, that is 
$p(s|s') \equiv p(\bm{l})$, $P(s|s';\xi)\equiv P(\bm{l};\xi)$, where 
$\bm{l} = s-s'$. It is known that 
$P(\bm{l};\xi) = (2\pi)^{-d}\int_B \mathrm{d} \bm{k}\, e^{-i\bm{k} \bm{l}}
\left[1-\xi\omega(\bm{k})\right]^{-1}$
where the integration is over the first Brillouin zone $B=[-\pi,\pi ]^d$ 
and $\omega(\bm{k}) = \sum_{\bm{l}} p(\bm{l})e^{i\bm{k} \bm{l}}$ is the 
structure function of the walk. While exact formulas are hard to obtain 
in $d \geq 4$, the leading order behavior of $\av{S_n}$ and $\av{X_n}$ can 
be extracted from applying the discrete Tauberian theorem 
\cite{RandWalksRandEnvVol1_H95} on the corresponding generating functions. 
According to this theorem, the scaling $c_n \sim n^{\kappa}$ as $n \gg 1$ is
equivalent to having the behavior $C(\xi) \simeq (1-\xi)^{-\kappa-1}L(1/(1-\xi))$
for the generating function $C(\xi) = \sum_{n=0}^{\infty} \xi^n c_n$ 
in the limit $\xi \to 1^-$. The results for $\av{S_n}$ are also summarized in  
\cite{RandWalksRandEnvVol1_H95}, we quote them here along with our results 
for $\av{X_n}$ for comparison and completeness. For $d=1$,  $\av{S_n} 
\sim \sqrt{8n/\pi}$ ,  $\av{X_n} \sim \sqrt{8n/\pi}$. For $d=2$, square lattice 
$\av{S_n} \sim \frac{\pi n}{\ln (8n)}$ ,  $\av{X_n} \sim 
\frac{4 \pi n}{3\pi+2 \ln(8n)}$ and for the triangular lattice 
$\av{S_n} \sim \frac{2\pi n}{\sqrt{3}\ln (12n)}$ , $\av{X_n} \sim  \frac{6 \pi n}
{5\pi + \sqrt{3} \ln{(12 n)}}$. For $d \ge 3$ cubic (hypercubic) lattices, simple 
random walks are non-recurrent (transient), hence $P(\bm{0};1^{-}) < \infty$ 
and we obtain $\av{S_n} \sim \frac{n}{P(\bm{0};1^{-})}$, $\av{X_n} 
\sim \frac{4dn}{2d-1+2 P(\bf{0};1^{-})}$ .  

\begin{figure}[htbp]
\centerline{\includegraphics[width=5.6in]{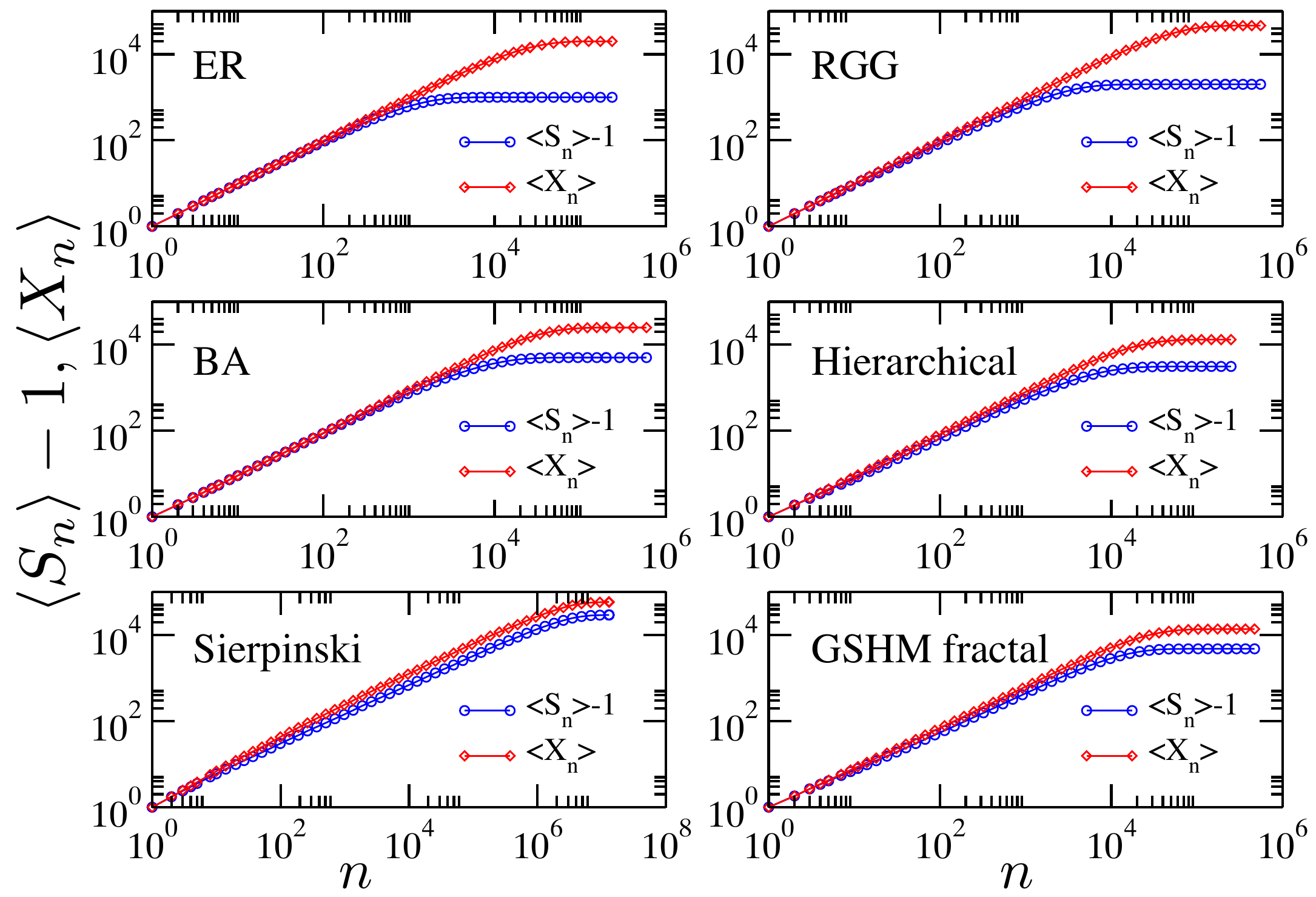}} 
\caption{Growth of $\av{S_n}-1$ 
(blue circles) and $\av{X_n}$ (red circles) for simple random walks. ER 
($N=1000$, $\av{k}=40$); RGG ($d=2$, $N=2000$, $\av{k}=50$); BA 
($N=5000$, $\av{k}=10$); Hierarchical network ($N=3125$, $\av{k}=8$); 
Sierpinski gasket ($N=29526$); GSHM fractal  ($N=4810$, $\av{k}=5.74$). 
Results were averaged for $300$ initial conditions on $200$ different 
graphs.} 
\label{fig:simple}
\end{figure}

These special cases suggest that for simple random walks $\lambda = \mu$, 
i.e., the edges are discovered mostly by visiting new nodes, and once
the nodes have all been visited, the remaining edges are discovered, at the
same rate. This holds for simple random walks on other networks
as well,  as indicated by our simulations
summarized in \reffig{simple}. 
Simulations were run on Erd\H{o}s-R\'enyi (ER) random graphs \cite{ErdosRgraph}, random 
geometric graphs (RGG) \cite{PREDall02_rgg} and the scale-free Barab\'asi-Albert (BA) model \cite{Science_BA99}, the 
hierarchical network model \cite{Ravasz_Hier02}, the fractal Sierpinski gasket and the GSHM fractal 
network \cite{PNASGallos07}. The curves for $\av{S_n}$ and $\av{X_n}$ almost
perfectly overlap, or run in parallel ($\lambda = \mu $). In the case of ER, RGG 
and BA models the growth rates are linear. For the hierarchical network 
$\lambda \simeq 0.94$, $\mu \simeq 0.99$, and for GSHM: $\lambda \simeq 0.92$ 
and $\mu \simeq 0.96$. For the Sierpinski gasket $\lambda \simeq 0.68$ 
(same as in \cite{Auriac_JPA_83}) and $\mu \simeq 0.73$. 
In all cases where deviations were observed for the exponents, they were small, 
on the order of $\mu - \lambda \leq 0.05$. One can show that these deviations 
are due to correction terms which, while vanish in the $N \to \infty$ limit, they 
still show up in the simulations (which are on relatively small networks, to be able 
to observe the saturations). It is possible to prove that
$\lambda = \mu$ in the $N \to \infty$ limit holds for general STP walks on arbitrary graphs, 
by  showing that $0 < b(1^{-})\overline{W}(s;1^{-}) < \infty$, see Appendix B.

\section{Adaptive walks: a simple bound}  

If STP walks are not good explorers, then naturally the question arises: What ATP walks
would have good discovery properties? Due to time dependence, ATP walks 
present a much wider array of possibilities and their systematic treatment is a hard 
problem.  Instead of tackling this general issue, here we first provide a simple {\em 
upper bound}   for the {\em mean} edge discovery growth exponent, 
obeyed by {\em any} walk  (ATP, or STP). Note that for ATP walks it is not necessarily 
true that $\av{S_n}$ or $\av{X_n}$ obeys scaling with a single exponent until saturation.
However, due to the constraints from \reffig{site_node_expl_pseudonode}a-b), the
`local slopes' still obey $1 \geq \mathrm{d} \av{X_n} / \mathrm{d} n \geq  \mathrm{d} \av{S_n} / \mathrm{d} n \geq 0$. 
Because slopes vary, we define the mean growth exponents 
$\mu = \ln M / \ln T_E$ and $\lambda = \ln N / \ln T_V$.
The bound is based on the observation 
that the edge cross-over time
cannot be smaller than the node cross-over time, $T_E \geq T_V$. This 
 provides the upper bound $\mu \leq \ln M / \ln (T_V)$. At $T_V$, however, 
$S_{n=T_V} \sim T_V^{\lambda} = N$, and thus $\ln (T_V) = \frac{1}{\lambda} \ln N$.
Recall that $M \sim N^{\nu}$.
Since the graph $G$ is connected, $1 \leq \nu \leq 2$.
We therefore find that:
\begin{equation}
\lambda \nu \geq \mu \geq \lambda\;.  \label{mmax}
\end{equation}
Hence a {\em necessary} condition for ATP walks to achieve  $\mu > \lambda$ 
mean growth exponents is $\nu > 1$.  If $\nu = 1$, 
(sparse graphs), clearly no walk (ATP or STP) can achieve $\mu > \lambda$.  
This is for example the case for all large networks with $N \to \infty$ that have  fixed 
maximum degree $D$, as
$M \leq D N$ and thus $\nu = 1$. 
Inequalities 
(\ref{mmax}) also show that the denser the graph, the larger the difference $\mu-\lambda$
could be, possibly obtained  by sufficiently ``smart" ATP walks. However, as $\nu \leq 2$, 
the mean edge discovery growth exponent can never be larger than twice that for nodes, 
i.e., $2 \lambda$.

\section{The Edge Explorer Model (EEM)} 

Next we introduce an ATP walk, the 
Edge Explorer Model, where the transition probability to step onto a neighboring 
site depends on the visitation history of that site and its neighbors. The EEM is 
one of the simplest models that performs enhanced graph discovery compared 
to STP walks, however, many other variants can be devised and fine tuned. 
\begin{figure}[htbp]
\centerline{\includegraphics[width=4.6in, height=2.5in]{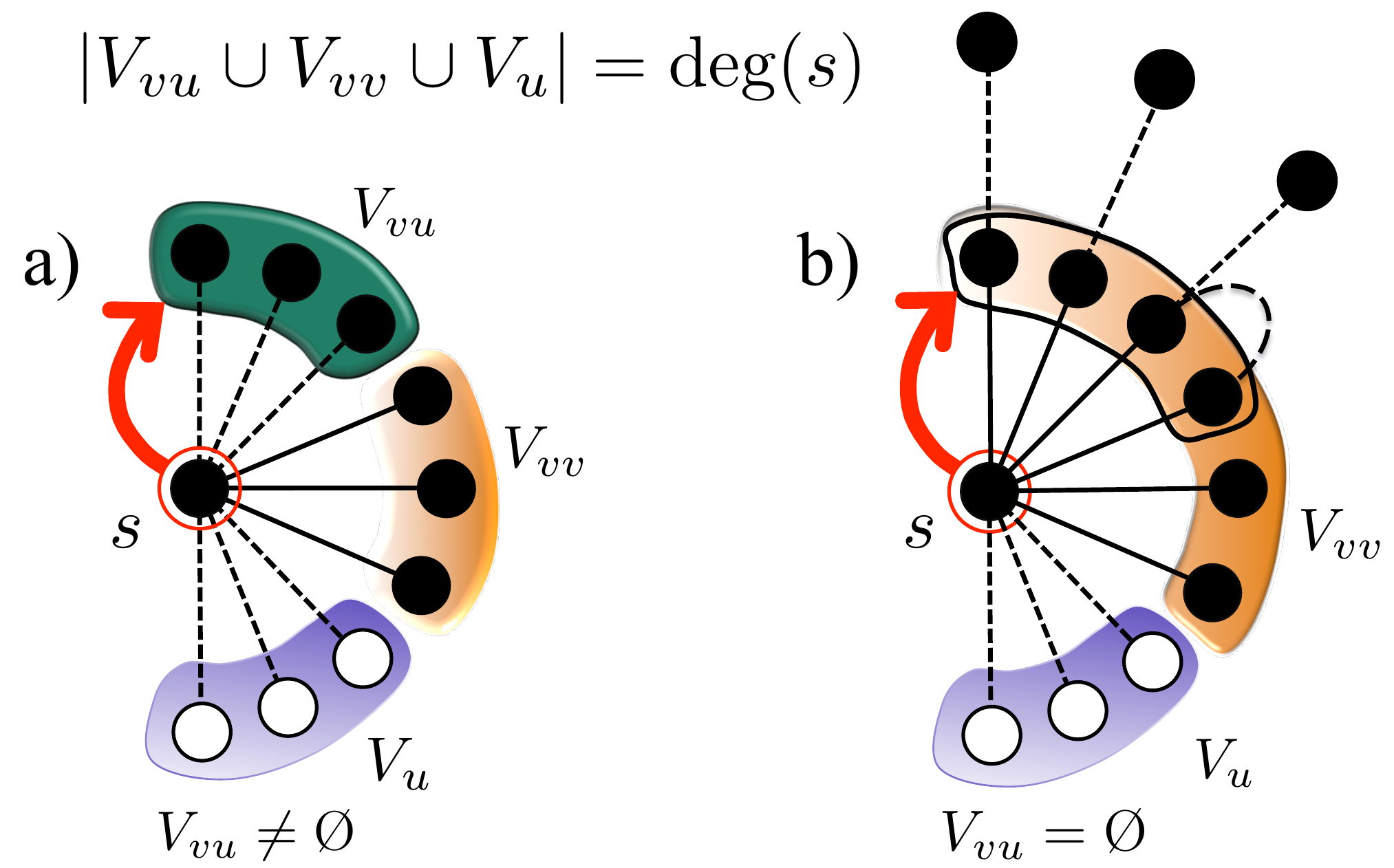}} 
\caption{The Edge Explorer Model. (a) If there are already discovered neighbors
$V_{vu}$  of site $s$ connecting through unvisited edges to $s$, the walker chooses
one at random, uniformly, to move to. 
(b) If the visited neighbors are connected to $s$ through visited edges only 
($V_{vu} = \emptyset$), the walker chooses one uniformly at random
from those that have an unvisited  connection to another visited node.}
\label{fig:adaptive_rw}
\end{figure} 

The immediate neighbors of a site $s$ can be divided into the set $V_{vv}$ 
of nodes that have already been visited and connected to $s$ via visited edges, 
the set of nodes $V_{vu}$ that have been already visited, but connected to 
$s$ via unvisited edges and the set $V_{u}$ of unvisited nodes (\reffig{adaptive_rw}).  
If $V_{vu} \neq \emptyset$ (\reffig{adaptive_rw} a)), the walker steps to one 
of the nodes {\em from this set},  chosen uniformly at random. 
If, however, $V_{vu} = \emptyset$ but $V_{vv} \neq \emptyset$ (\reffig{adaptive_rw} b)), 
the walker chooses a node uniformly at random among the nodes
within $V_{vv}$ that have at least one connection via 
{\em unvisited} edges to other visited nodes. If no such nodes exist, then
the walker chooses a node uniformly at random from $V_u$.
\begin{figure}[htb]
\centerline{\includegraphics[width=4.8in]{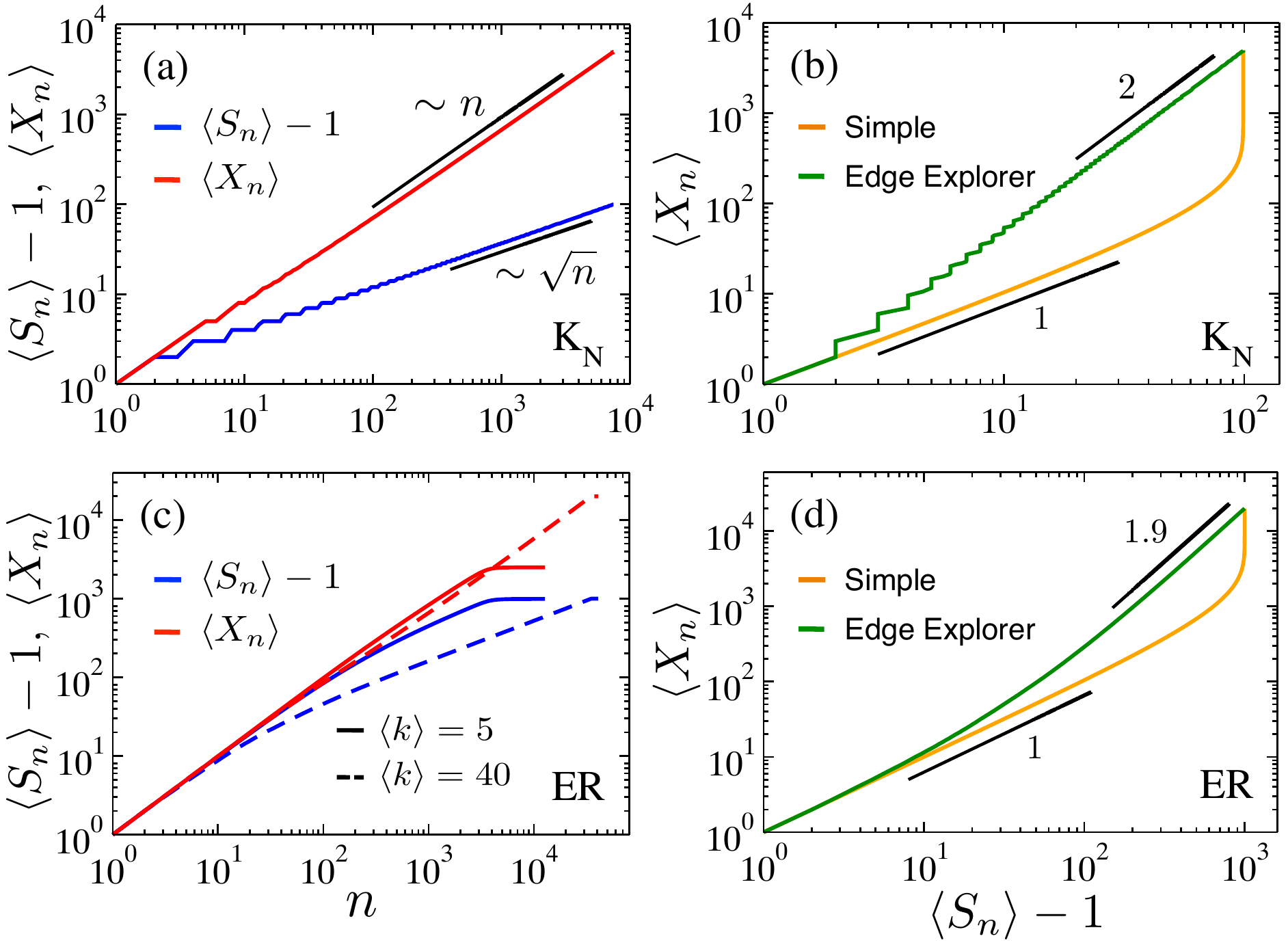}} 
\caption{Site and edge exploration growth curves obtained for EEM
on complete graphs ( $K_N$, $N=100$, a) and b)), Erd\H{os}-R\'enyi graphs 
(ER, $N=1000$, c) and d)). }
\label{fig:sixfigures}
\end{figure} 
While in general ATP walks do not lend to analytical treatment, all the
properties of the EEM model on the complete graph $K_N$ can be obtained
exactly. Due to its rules, the walker always discovers {\em at least one edge} 
in two steps (only true on $K_N$ !), which means that edge exploration 
happens linearly in time,  $\mu = 1$. Assuming that it has discovered 
$m-1$ nodes, it will {\em not} discover a new node until it has discovered all the
links amongst the $m-1$ nodes, the discovered graph becoming $K_{m-1}$. 
Then it adds the $m$-th node, discovering the remaining $m-1$ edges  in 
${\cal O}(m)$ steps, thus finishing discovering all the nodes in 
${\cal O}(\sum m) = {\cal O}(N^2)$ steps.   This means $T_V = {\cal O}(N^2)$
and therefore $\lambda = \ln N/\ln T_V = 1/2$. Since on $K_N$ the
EEM walker cannot get lost in visited regions, the corresponding cross-over 
times and cover times are practically  the same. 
\reffig{sixfigures}a) shows  $\av{S_n}$ and $\av{X_n}$ with these predicted 
features, including $\mu=2\lambda = 1$. On $K_N$, the edge exploration is 
optimal in the sense that for a given number of discovered nodes, it discovers
the maximum number of edges possible up to that point, as shown in
\reffig{sixfigures}b). \reffig{sixfigures}c-d) show the same for EEM on ER 
graphs. As discussed above, the $\mu -\lambda$ slope difference increases 
with graph density, defined as $\rho = \rho(G) =2M/[N(N-1)] \leq 1$. On sparse 
graphs, however, the EEM can get trapped in visited regions if these regions
are clusters/communities separated by bottlenecks from the rest of the graph. 
Within these regions the EEM walker performs a simple random walk
before it escapes. For this reason, 
on low-density graphs the EEM is not necessarily the optimal explorer. 
To illustrate the graph discovering fidelity of the EEM, in \reffig{densities} we 
compare the densities $\overline{\rho}_n = 2 \av{X_n}/\av{S_n} (\av{S_n}-1)$ of the 
discovered graphs as function of time $n$, generated on the same network 
by the EEM and by the  simple random walk (for $K_N$ and ER). 
\begin{figure}[hbt]
\centerline{\includegraphics[width=5.3in]{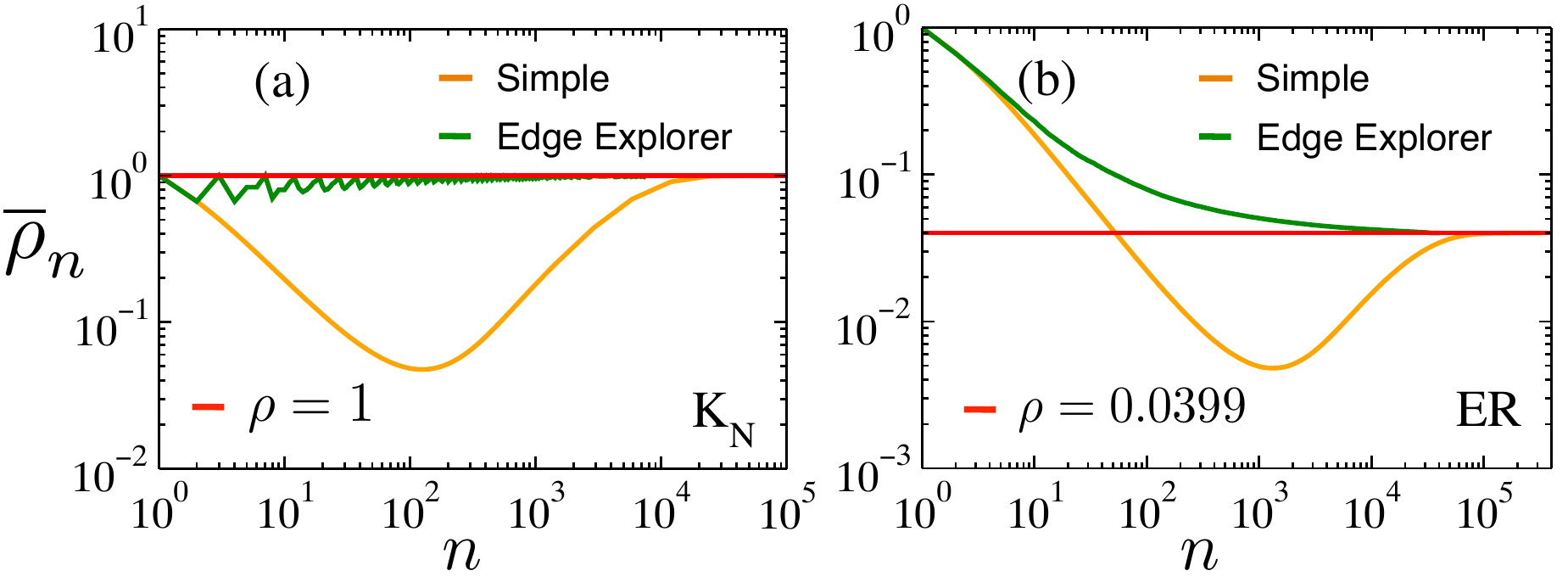}} 
\caption{Comparison of the performance in graph discovery by the EEM adaptive
walk (green) and by the simple random walk (orange), on a) $K_N$ and 
b) ER graphs, by plotting the  time evolution of the discovered graph's density. 
The red line is the true density. For a) $N = 100$, for b) $N = 10^3$, 
$\langle k \rangle = 40$. }
\label{fig:densities}
\end{figure} 
Clearly, the simple random walk greatly undershoots the  true graph
density (indicated by horizontal red line), corresponding to 
$\av{X_n}  = {\cal O}(\av{S_n})$, shown earlier, before it starts closing on the
true value $\rho(G)$; on the contrary, the EEM shows a systematic and 
rapid approach to $\rho(G)$. 

\section{Discussion} 
In summary, we have investigated properties of network discovery by
walkers that follow edges (also called crawlers) in the most general setting. 
We have derived an exact expression for the average number of discovered
edges $\av{X_n}$ (its generating function) as function of time for arbitrary
graphs and STP walks.
In particular, we have shown that for STP walkers both edge and node 
discovery follow the same scaling law on large networks, independently on 
the form of the stationary transition probabilities.  Hence, the discovered
network will be sparse (the number of discovered edges scaling linearly
with that of the discovered nodes), presenting a strongly skewed 
structure compared to the underlying network's if the latter is
not sparse, $\nu > 1$. Only after a cross-over time $\sim {\cal O}(N)$, 
will the edges become increasingly discovered, which in the case of large 
networks means unfeasibly large wait times, eliminating STP walks 
as a useful methodology for faithful network discovery.  Our results thus
rigorously show that efficient/faithful discovery can only be done with 
adaptive walkers, whom use time/history dependent information for their 
transition probabilities (ATP). Visiting history information can be thought of
as ``pheromone'' trails on the network, which the walker uses through its
rules for stepping onto the next site \cite{Koenig01ants}. There is a plethora 
of possible rules
using past history, however, to keep memory requirements low (bounded) 
on a walker, the desirable rules are the ones that only use information from
the local neighborhood of the walker. In this vein, we have introduced
a simplistic  adaptive walk, the Edge Explorer Model, which is greedily biased
towards already visited regions within a 2-step neighborhood. We have shown
that on dense graphs the EEM performs near optimally  or optimally (on $K_n$).\\

\medskip

This project was supported in part by the Army Research Laboratory, ARL Cooperative Agreement Number W911NF-09-2-0053,  HDTRA 201473-35045 and NSF 
BCS-0826958.
The content of this document are those of the authors and should not be interpreted as 
representing the official policies, either expressed or implied, of the ARL or the U.S. Government.
The authors thank R.K.P Zia, 
B. Szymanski, M. Ercsey-Ravasz and S. Sreenivasan for useful discussions.


\appendix
\section{Derivation of $X(s_0,\xi)$}

  Consider a connected simple graph $G(V,E)$ (there is a path 
along $G$'s edges between any two nodes), where $V$
denotes the set of nodes and $E$ the set of edges.
In these pages we show the details of calculations for various 
exploration properties of (general)
random walks on $G$, the only constraint being that the walk
is restricted to move along existing edges (no long-range hops
are allowed).  We will rely heavily on the standard 
generating function technique \cite{RandWalksRandEnvVol1_H95,WilfGfunction05} and for that reason 
we briefly introduce related definitions and basic results.
In particular, if $A_j$ is an arbitrary series, then its generating function $A(\xi)$ with 
$|\xi| < 1$ is defined as:
\begin{equation}
A(\xi) = \sum_{n=0}^{\infty} \xi^{n} A_n 
\label{eq:002}.
\end{equation}
Knowing  $A(\xi)$,  the elements of the series $A_n$ are 
recovered by inverting (\ref{eq:002}):
\begin{equation}
A_n=\frac{1}{2\pi i} \oint\limits_{\Gamma} \frac{d\xi}{\xi^{n+1}} A(\xi)\;,
\label{eq:003}
\end{equation}
where $\Gamma$ is counterclockwise contour around $\xi=0$. We are going to make 
use of the following expressions: 
\begin{equation}
\sum_{j=0}^{n} A_j = \frac{1}{2\pi i} \oint\limits_{\Gamma} 
\frac{d\xi}{\xi^{n+1}} \frac{1}{1-\xi}A(\xi)\;,\;\;\; \mbox{and}\;\;\;
\sum_{j=0}^{\infty} A_j  = \lim_{\xi \to 1, \atop \xi < 1} A(\xi) \equiv A(1^-)\;. \label{eq:004}
\end{equation}
On many occasions, the inversion integral in (\ref{eq:003}) cannot be performed analytically. 
However, we are usually interested in the long-time limit $n \gg 1$ of the quantities, and for that
we use  the discrete Tauberian theorem which allows to estimate the leading order term,
or the Darboux theorem (which can also produce terms beyond the leading order) 
\cite{RandWalksRandEnvVol1_H95,IntrProbTheorApplVol2_F71}.

\medskip

  Let $P_n(s|s_0)$ denote the probability of the walker being at site $s$ on the $n$th step
given that it started from site $s_0$, and let $F_n(s|s_0)$ denote the probability of the
walker visiting site $s$ for {\em the first time} on the $n$-th step, given it started from $s_0$. 
The corresponding generating functions are $P(s|s_0;\xi)$ and $F(s|s_0;\xi)$ (Note that 
$F(s|s_0;\xi) = \sum_{n=1}^{\infty} \xi^{n} F_n(s|s_0)$). Partition over the last step and 
partition over the first step give two useful recursion relations ($r,s_0 \in V $):
\begin{equation}
P_{n+1}(r|s_0)=\sum_{r'\in V} p(r|r')
P_{n}(r'|s_0),\;\;\;\;\;\;P_{n+1}(r|s_0)=\sum_{r'\in V} P_{n}(r|r')
p(r'|s_0)\;.\label{pols}
\end{equation}
In terms of generating functions:
\begin{eqnarray}
&&P(r|s_0;\xi)=\delta_{rs_0}+\xi\sum_{r'\in V} p(r|r')
P(r'|s_0;\xi)\;,\label{posx}\\
&&P(r|s_0;\xi)=\delta_{rs_0}+
\xi\sum_{r'\in V} P(r|r';\xi)p(r'|s_0)\;.\label{polx}
\end{eqnarray}
These identities can be used to derive a relationship between the site occupancy 
generating function and the first-passage time generating function, valid for {\em all} 
connected graphs, and {\em all}  walks \cite{IntrProbTheorApplVol2_F71}:
\begin{equation}
F(s|s_0;\xi) = \frac{P(s|s_0;\xi)-\delta_{ss_0}}{P(s|s;\xi)}\;. \label{eq:005}
\end{equation}
\vspace*{0.5cm}


\medskip

  In the following we derive Eqs (5), (6) of the main paper. 
Let $e=(s,s')\in E$ be an edge in  $G$, and let $F_n(e|s_0)$ denote the 
probability that the walker passes through $e$ {\em for the first time} on the 
$n$-th step, given it commenced from  $s_0 \in V$. Let us denote by $X_n(s_0)$ 
the number of distinct edges visited during an $n$-step walk that 
commenced from site $s_0$. We introduce the indicator $\Gamma_n(s_0) \in \{0,1\}$ 
for the number of virgin edges discovered on the $n$-th step: 
\begin{equation}
\langle \Gamma_n (s_0)\rangle = \mbox{Prob}\{ \Gamma_n(s_0) = 1\} = 
\sum_{e\in E} F_n(e | s_0) \;.
\label{eq:010}
\end{equation}
As convention we take $\Gamma_0 = \langle \Gamma_0 \rangle=0$. Thus we can write:
\begin{equation}
X_n(s_0) = \sum_{j=0}^n \Gamma_j(s_0)\;,\;\;\;\langle X_n(s_0) \rangle = \sum_{j=0}^{n} 
\langle \Gamma_j(s_0) \rangle\;,
\label{eq:011}
\end{equation}
The corresponding generating function is :
\begin{equation}
X(s_0;\xi) = \sum_{n=0}^{\infty} \xi^n \langle X_n(s_0) \rangle = 
\sum_{n=0}^{\infty} \xi^n \sum_{j=0}^{n} \langle \Gamma_j(s_0) \rangle
=\frac{\Gamma(s_0;\xi)}{1-\xi}\;,
\label{eq:012}
\end{equation}
where $\Gamma(s_0;\xi)$ is the generating function for the indicator and we used the first
identity in (\ref{eq:004}). 
From (\ref{eq:010}):
\begin{equation}
\Gamma(s_0;\xi) = \sum_{e\in E} \sum_{n=1}^{\infty} \xi^n F_n(e|s_0)=
\sum_{e\in E} F(e|s_0;\xi). \label{eq:013}
\end{equation}
We need to calculate the edge first passage probabilities $F_n(e|s_0)$, or their generating
function $F(e|s_0;\xi)$. Clearly, $F_n(e|s_0)$ contains all the paths commencing from 
$s_0$ that {\em never crossed} edge $e$ during the first $n-1$ steps, but they do so on the
$n$-th step. 

To calculate $F(e|s_0;\xi)$, we  introduce an auxiliary node $z$ placed on the 
edge $e$, as described in the main
paper (Fig. 1c)), and consider the random walk on this extended graph 
$G_z$. 
The node set of this graph is $V_z = V\cup \{z\}$ and the edge set 
$E_z = E\setminus \{(s,s')\}\cup \{(s,z),(z,s')\}$. 
 Let $P^{\dagger}_n(r|s_1)$ be the site occupation probability  and 
 $F^{\dagger}_n(z|s_1)$ be  the corresponding first passage time 
 distribution   for the random walk on $G_z$, and here $r,s_1 \in V_z$. 
Then it certainly holds that the first passage probability through edge 
$e$ on the $G$ graph is identical to the {\em site} first passage probability of 
the new walk to site $z$ on the $G_z$ graph:
\begin{equation}
F_n(e|s_0) = F^{\dagger}_n(z|s_0)\;,\;\;\;\;s_0\in V\;.
\label{eq:016}
\end{equation}
The corresponding generating function takes the form:
\begin{equation}
F(e|s_0;\xi) = F^{\dagger}(z|s_0;\xi)=
\frac{P^{\dagger}(z|s_0;\xi)}{P^{\dagger}(z|z;\xi)}\;,\;\;\; s_0 \neq z\;. 
\label{eq:017} 
\end{equation}
where we used the general identity (\ref{eq:005}) for the new walk on the new graph $G_z$.

In order to fully specify the new walk on $G_z$ we need to define the corresponding 
single-step transition probabilities. The single-step probabilities 
away from the nodes $s$ and $s'$  of the new walk are identical to the old walk's, including
those of getting to  $s$ and to $s'$ from  nodes other than $s'$ and $s$, respectively. 
When considering steps from 
one ($s$ ($s'$)) to the other ($s'$ ($s$)), the single-step probabilities in the new walk are forbidden. 
Instead the walker has to step onto site $z$ first, before it can step further to 
the other site ($s$ or $s'$). We also have to make sure that we have the same 
{\em probability flow} in the new walk as in the old one
when moving from  nodes $s$ and $s'$ towards $z$ (in the old walk
these would be along edge $e$). From node $z$ the walker can only step to $s$ or 
$s'$ with probabilities $f$ and $g$ respectively. 
The probabilities $f$ and $g$ are arbitrary, and the independence of the
final expression for $F(e|s_0;\xi)$ on these two variables serve as a  good 
test for the correctness of our calculations. Eq (4) of the main paper 
provides the condensed form of the  single-step transition probabilities for the new walk 
on $G_z$, and it combines the following cases:
\begin{eqnarray}
&&\!\!p^{\dagger}(s|s') = p^{\dagger}(s'|s)=0\;\;\;\; \label{pd1}  \\
&&\!\!p^{\dagger}(s|z) = f\;,\;\;\; p^{\dagger}(s'|z)=g ,\;\;\;
p^{\dagger}(r|z)=0\;\;\mbox{for} \;\;\; r\in V_z\setminus \{s,s'\} \;\;\;\;  \label{pd2} \\
&&\!\!p^{\dagger}(z|s')=p(s|s'),\;\;\;p^{\dagger}(z|s)=p(s'|s),\;\;
p^{\dagger}(z|r')=0\;\;\mbox{for} \;\;\; r'\in V_z\setminus \{s,s'\}\;\;\;\; \label{pd3} \\
&&\!\!p^{\dagger}(r|r')=p(r|r')+q(r|r'),\;\;q(r|r')=-(\delta_{r's'}\delta_{rs} + 
\delta_{r's}\delta_{rs'})p(r|r'),\;\;r,r'\in V.\;\;\;\; \qquad \label{eq:015}
\end{eqnarray}
The pseudo-node $z$ uniquely characterizes the 
edge $e=(s,s')\in E$. To remind us about this identification, we will write $z$ instead of $e$, 
in the remainder. 
From (\ref{eq:017}) and (\ref{eq:013}) it follows that:
\begin{equation}
\Gamma(s_0;\xi)=\sum_{z \in E} \frac{P^{\dagger}(z|s_0;\xi)}
{P^{\dagger}(z|z;\xi)}\;,\;\;\;\; s_0 \in V \;.
\label{eq:018}
\end{equation}
The sum is over all the edges of the {\em original} graph $G$. 
Thus we need to compute the site occupation probabilities for 
the new walk on the extended graph. 
The relationships based on partition over the last step  (\ref{posx}) and first 
step (\ref{polx}), hold for the new walk as well:
\begin{eqnarray}
&&P^{\dagger}(r|s_1;\xi)=\delta_{rs_1}+\xi\sum_{r'\in V_z} p^{\dagger}(r|r')
P^{\dagger}(r'|s_1;\xi)\;,\;\;\;r,s_1 \in V_z  \label{dposx} \\
&&P^{\dagger}(r|s_1;\xi)=\delta_{rs_1}+\xi\sum_{r'\in V_z} P^{\dagger}(r|r';\xi)
p^{\dagger}(r'|s_1)\;,\;\;\;r,s_1 \in V_z\;.\label{dpolx}
\end{eqnarray}
From (\ref{dposx}) with $r=z$ and $s_1=s_0$ and using (\ref{pd3}) one obtains:
\begin{equation}
P^{\dagger}(z|s_0;\xi)=\xi p(s|s') P^{\dagger}(s'|s_0;\xi) + \xi p(s'|s)
P^{\dagger}(s|s_0;\xi)\;,\;\;\;s_0\in V\;. \label{eq:023}
\end{equation}
Eq (\ref{dpolx}) with $r=s_1=z$ and (\ref{pd2}) yields:
\begin{equation}
P^{\dagger}(z|z;\xi)=1+\xi f P^{\dagger}(z|s;\xi) + \xi g P^{\dagger}(z|s';\xi)\;. \label{eq:024}
 \end{equation}
 Since (\ref{eq:023}) is valid for any $s_0 \in V$, we first make $s_0 \mapsto s$ and then 
the  $s_0 \mapsto s'$ substitutions to get :
 \begin{eqnarray}
&&P^{\dagger}(z|s;\xi)=\xi p(s|s') P^{\dagger}(s'|s;\xi) + \xi p(s'|s)
 P^{\dagger}(s|s;\xi) \label{eq:025} \\
 &&P^{\dagger}(z|s';\xi)=\xi p(s|s') P^{\dagger}(s'|s';\xi) + \xi p(s'|s)
 P^{\dagger}(s|s';\xi) \label{eq:026}
 \end{eqnarray}
Inserting these into (\ref{eq:024}) one obtains: 
 \begin{eqnarray}
P^{\dagger}(z|z;\xi) = 1+\xi^2p(s|s')\left[ f P^{\dagger}(s'|s;\xi)
+g P^{\dagger}(s'|s';\xi)  \right]&& +  \nonumber \\
+ \xi^2 p(s'|s) \left[ f P^{\dagger}(s|s;\xi) + g P^{\dagger}(s|s';\xi)  \right] && \label{eq:027}
 \end{eqnarray}
Eq (\ref{eq:027}) shows that we need to calculate the site 
occupation probability generating function $P^{\dagger}(r|s_0;\xi)$ for $r,s_0 \in V$.
In order to derive $P^{\dagger}(r|s_0;\xi)$ 
for $r,s_0 \in V$ we first make the replacements $r' \mapsto r''$,  $r \mapsto r'$, 
$s_1 \mapsto s_0 \in V$ in (\ref{dpolx}) to obtain:
\beq{b01}
P^{\dagger}(r'|s_0; \xi) =\delta_{r's_0} +\xi \sum_{r'' \in V_z} p^{\dagger} (r'|r'') P^{\dagger} (r''|s_0;\xi)\;,
\eeq
With the help of  (\ref{eq:015}) we then find:
$$
P^{\dagger}(r'|s_0; \xi) =\delta_{r's_0} +\xi \sum_{r'' \in V}\{ p(r' |r'') + q(r'|r'')\} P^{\dagger} (r''|s_0;\xi) + \xi p^{\dagger}(r'|z;\xi) P^{\dagger}(z|s_0;\xi)
$$
or after rearranging
\beqar{b03}
P^{\dagger}(r'|s_0; \xi) - \xi \sum_{r'' \in V} p(r'|r'') P^{\dagger} (r''|s_0;\xi) &=& \delta_{r's_0}
+ \xi \sum_{r'' \in V} q(r'|r'') P^{\dagger} (r''|s_0;\xi) \nonumber \\
 &+& \xi p^{\dagger}(r'|z;\xi) P^{\dagger}(z|s_0;\xi)\;.
\eeqar
After multiplying both sides with $P(r|r';\xi)$, $r \in V$ and summing both sides over $r' \in V$, the above equation takes the form of
\beqar{b04}
\sum_{r'\in V} P(r|r';\xi) P^{\dagger}(r'|s_0; \xi)&-& \sum_{r'' \in V}  P^{\dagger} (r''|s_0;\xi)\; 
\xi \sum_{r'\in V} p(r'|r'') P(r|r';\xi) =P(r|s_0;\xi) + \nonumber \\
 &+&\sum_{r'' \in V} P^{\dagger}(r''|s_0; \xi) \; \xi \sum_{r' \in V} P(r|r';\xi)q(r'|r'') + \nonumber \\
 &+& \xi [f P(r|s;\xi)  + g P(r|s';\xi)] P^{\dagger}(z|s_0;\xi).
\eeqar 
Next we calculate the left hand side  of (\ref{eq:b04}) by using (\ref{polx}) to write: 
$$
\xi \sum_{r'\in V} p(r'|r'') P(r|r';\xi) = P(r|r'';\xi) - \delta_{rr''}
$$ 
When this is inserted into the lhs of (\ref{eq:b04}), the sums with dagger terms cancel and one
just simply obtains $P^{\dagger} (r|s_0;\xi)$. 
The sums on the right hand side can be written in a simpler form after 
introducing the notation:
\beq{b05}
A(r|r'' ;\xi) =\xi \sum_{r' \in V} P(r|r';\xi) q(r'|r'').
\eeq
Thus, eq (\ref{eq:b04}) assumes the expression:
\begin{eqnarray}
P^{\dagger}(r|s_0; \xi) = P(r|s_0;\xi) &+& \xi \left[ f P(r|s;\xi) + g P(r|s';\xi)\right]
P^{\dagger}(z|s_0;\xi) \nonumber \\
&+&\sum_{r'' \in V} A(r|r''; \xi) P^{\dagger}(r''|s_0;\xi)\;. \label{PA}
\end{eqnarray}
Using (\ref{eq:015}) we find:
\begin{equation}
A(r|r''; \xi)  = - \xi P(r|s;\xi) p(s|r'')\delta_{r''s'} - \xi   P(r|s';\xi) p(s'|r'')\delta_{r''s}
\label{eq:030}
\end{equation}
Inserting this in (\ref{PA}) one obtains:
\begin{eqnarray}
P^{\dagger}(r|s_0;\xi) &= &P(r|s_0;\xi) -  \xi P(r|s;\xi) p(s|s')P^{\dagger}(s'|s_0;\xi)- 
\xi   P(r|s';\xi) p(s'|s)P^{\dagger}(s|s_0;\xi) + \nonumber \\
&+&\xi \left[ f P(r|s;\xi) + g P(r|s';\xi)\right]P^{\dagger}(z|s_0;\xi)\;,\;\;\;r \in V\;. 
\label{eq:031}
\end{eqnarray}
Replacing $r \mapsto s$ and then $r \mapsto s'$ in the equation above,  yields:
\begin{eqnarray}
\left\{
\begin{array}{l}
a_{11} P^{\dagger}(s|s_0;\xi) + a_{12} P^{\dagger}(s'|s_0;\xi) + a_{13} P^{\dagger}(z|s_0;\xi) = 
P(s|s_0;\xi)  \\
a_{21} P^{\dagger}(s|s_0;\xi) + a_{22} P^{\dagger}(s'|s_0;\xi) + a_{23} P^{\dagger}(z|s_0;\xi) = 
P(s'|s_0;\xi) 
\end{array}
\right.  \label{eq:032}
\end{eqnarray}
where:
\begin{eqnarray}
\begin{array}{ll}
a_{11} = 1+\xi p(s'|s) P(s|s';\xi) \qquad & a_{21} = \xi p(s'|s) P(s'|s';\xi) \\
a_{12} = \xi p(s|s') P(s|s;\xi)  \qquad & a_{22} = 1+\xi p(s|s') P(s'|s;\xi) \\
a_{13} = - \xi \left[ f P(s|s;\xi) + g P(s|s';\xi)\right]  \qquad & a_{23} = 
- \xi \left[ f P(s'|s;\xi) + g P(s'|s';\xi)\right]
\end{array}
\label{eq:033}
\end{eqnarray}
System (\ref{eq:032}) needs a third equation, to solve for $\{ P^{\dagger}(s|s_0;\xi), 
P^{\dagger}(s'|s_0;\xi), P^{\dagger}(z|s_0;\xi) \}$. The third
equation is just (\ref{eq:023}):
\begin{equation}
a_{31} P^{\dagger}(s|s_0;\xi) + a_{32} P^{\dagger}(s'|s_0;\xi) + a_{33} P^{\dagger}(z|s_0;\xi) = 
0 
\label{eq:034}
\end{equation}
with:
\begin{equation}
a_{31} = \xi p(s'|s)\;,\;\;\;\;a_{32} = \xi p(s|s')\;,\;\;\;\;a_{33} = -1 
\label{eq:035}
\end{equation}
Thus, if we introduce the column vectors:
\begin{eqnarray}
\left[ P^{\dagger} \right] (s,s',z|s_0;\xi)  \equiv
\left[
\begin{array}{c}
P^{\dagger}(s|s_0;\xi) \\
P^{\dagger}(s'|s_0;\xi) \\
P^{\dagger}(z|s_0;\xi) 
\end{array}
\right] \;, \;\;\;\;\;\;\;
\left[ P\right] (s,s'|s_0;\xi)  \equiv
\left[
\begin{array}{c}
P(s|s_0;\xi) \\
P(s'|s_0;\xi) \\
0
\end{array} \right] \;, \label{eq:037}
\end{eqnarray}
and denote by ${\bf A}$ the $3\times3$ matrix with 
elements defined above by (\ref{eq:033}) and (\ref{eq:035}), 
then the linear equation to be solved is simply:
\begin{equation}
{\bf A} \left[ P^{\dagger} \right] (s,s',z|s_0;\xi) = \left[ P\right] (s,s'|s_0;\xi)\;.  \label{eq:038}
\end{equation}
Assuming that  ${\bf A}$ is invertible, the solution is
\begin{equation}
\left[ P^{\dagger} \right] (s,s',z|s_0;\xi) = {\bf A}^{-1}\left[ P\right] (s,s'|s_0;\xi) 
\label{eq:039}
\end{equation}
The matrix explicitly looks as follows:
\begin{eqnarray}
{\bf A} = \left[ 
\begin{array}{ccc}
1+\xi \alpha a & \xi \beta b & -\xi (g a+ f b) \\
\xi \alpha c & 1+ \xi \beta d & -\xi (g c+ f d) \\
\xi \alpha & \xi \beta & -1 
\end{array}
\right]
\label{eq:040}
\end{eqnarray}
where we introduced the shorthand notations:
\begin{eqnarray}
&&\alpha = p(s'|s)\;,\;\;\; \beta = p(s|s')\;, \label{eq:041a}\\
&&a=a(\xi) = P(s|s';\xi)\;,\;\;\; b = b(\xi) = P(s|s;\xi)\;,\;\;\;  \label{eq:041b} \\
&&c = c(\xi) =P(s'|s';\xi)\;,\;\;\; d =d(\xi) =P(s'|s;\xi)  \label{eq:041c}
\end{eqnarray}
The inverse of  matrix ${\bf A}$ is just
\begin{eqnarray}
{\bf A^{-1}} = \frac{1}{D}\left[
   \begin{array}{ccc}
   -1- \beta(d -v  \xi)\xi & \beta(b- u \xi) \xi  &  (u +t \beta g \xi) \xi \\
   \alpha(c-v \xi) \xi & -1- \alpha(a - u \xi) \xi & (v+t\alpha f \xi)\xi \\
   - \alpha (1+\bar{v} \beta \xi) \xi & -\beta(1+\bar{u}\alpha \xi)\xi & 1+(w +t \alpha \beta \xi)\xi
   \end{array}
\right] 
\label{eq:042}  
\end{eqnarray}
where the determinant is:
 \begin{equation}
D= -1- w\xi + (u\alpha +v\beta -t\alpha \beta) \xi^{2}+ t \alpha \beta (f + g) \xi^{3},\label{eq:043}
 \end{equation} 
and we introduced the notations:
\begin{eqnarray}
u=g a+f b \quad & \bar{u} = a-b \nonumber \\
v=f d+ g c \quad & \bar{v} =d-c \label{eq:044}\\
t=ad-bc \quad & w=a\alpha+d\beta. \nonumber
\end{eqnarray} 
The final solutions  can be easily read from (\ref{eq:039}). Note that these are now expressed solely in terms
of the generating functions for the site occupation probabilities of the
old walk on the old graph $G$! In particular for $ P^{\dagger}(z|s_0;\xi)$ we get:
\begin{equation}
P^{\dagger}(z|s_0;\xi)= -\frac{\alpha \xi}{D} (1+\bar{v}\beta \xi) P(s|s_{0};\xi) -\frac{\beta \xi}{D}(1+\bar{u} \alpha \xi) P(s'|s_{0};\xi). \label{eq:045}
\end{equation}
In order to obtain the site occupation probability generating functions $ P^{\dagger}(s|s;\xi)$, 
$ P^{\dagger}(s'|s;\xi)$, $ P^{\dagger}(s|s';\xi)$ and $ P^{\dagger}(s'|s';\xi)$ as required by 
the r.h.s. of (\ref{eq:027})
we merely substitute $s$ and $s'$, respectively for $s_0$ in the expressions for the solutions. 
We thus obtain:
 \begin{equation}
 P^{\dagger}(s|s;\xi)=-\frac{1}{D}(b+ g \beta t\xi^2)\;,\;\;\;\;\;
 P^{\dagger}(s|s';\xi)=-\frac{1}{D} (a + t\beta\xi-t\beta f \xi^2)  \label{eq:049}
 \end{equation}
\begin{equation}
 P^{\dagger}(s'|s;\xi)=-\frac{1}{D} (d+t\alpha \xi -t\alpha g \xi^2)\;,\;\;\;\;\;
 P^{\dagger}(s'|s';\xi)=-\frac{1}{D} (c+t\alpha f \xi^2).  \label{eq:051}
 \end{equation}
Inserting these into (\ref{eq:027}):
\begin{equation}
P^{\dagger}(z|z;\xi)=-\frac{1}{D}(1+w\xi+t\alpha \beta \xi^2)\;. \label{eq:052}
\end{equation}
Next, from Eqs. (\ref{eq:045}) (\ref{eq:052}) and (\ref{eq:018}) we find:
\begin{equation}
\Gamma(s_0;\xi)=\!\!\!\sum_{z \in E} \frac{\alpha \xi [1+(d-c) \beta \xi] P(s|s_{0};\xi) + \beta \xi [1+(a-b)\alpha \xi] P(s'|s_{0};\xi)}{1+(a \alpha + d\beta)\xi + (ad-bc) \alpha \beta \xi^{2}}. \label{eq:053}
\end{equation} 
Note that this expression is independent on the variables $f$ and $g$, as anticipated!
Now using (\ref{eq:012}), the generating function for the average number of discovered distinct 
edges becomes:
\begin{equation}
X(\xi)=\frac{\xi}{2(1-\xi)}  \sum_{s,s'}\frac{\alpha [1+(d-c) \beta \xi] P(s|s_{0};\xi) + \beta [1+(a-b)\alpha \xi] P(s'|s_{0};\xi)}{1+(a \alpha + d\beta)\xi + (ad-bc) \alpha \beta \xi^{2}}.
\label{eq:054}
\end{equation}
Taking a closer look at this expression one observes that only those $(s,s')$ pairs will contribute in the sum which are neighbors on $G$ (since $\alpha$ and $\beta$ are zero for transitions along non-edges). After adding the sum to itself then interchanging the  dummy variables $s, s'$ in one of the sums, we finally obtain eqs (5),(6) of the main paper:
\begin{equation}
X(s_0;\xi)=\frac{\xi}{1-\xi} \sum_{s} \overline{W}(s;\xi) P(s|s_{0};\xi)\;, \label{eq:056}
\end{equation}
with
\begin{equation}
\overline{W}(s;\xi)=\sum_{s'}\alpha \frac{1+(d-c)\beta \xi}{1+(a 
\alpha +d \beta)\xi + (ad-bc) \alpha \beta \xi^{2}} \;.\label{eq:057}
\end{equation}

\section{Properties of $X(s_0;\xi)$}

  Next we show a number of properties of the weight function $\overline{W}(s;\xi)$. Before
we do that, however, we need to establish a number of fundamental inequalities involving
the generating functions (\ref{eq:041b}), (\ref{eq:041c}).  Let us use the temporary notation
(where all quantities are understood implicitly in the $\xi \to 1^{-}$ limit, $\xi \in \mathbb{R}$):
\begin{equation}
\Delta \equiv 1+ a\alpha + d\beta + (ad-bc) \alpha \beta  \label{ddenom}
\end{equation}
The denominator $\Delta$ of (\ref{eq:057}) appears in the numerator of $P^{\dagger}(z|z;\xi)$ given by (\ref{eq:052}) and thus:
\begin{equation}
P^{\dagger}(z|z;\xi) = -\dfrac{\Delta}{D} \;. \label{pidagdelta}
\end{equation}
Since the $P^{\dagger}$-s in (\ref{eq:049}-\ref{eq:052}) are all generating functions for
random walk probabilities, (that is, they are power-series with positive  coefficients) they are
all positive (for $\xi \in \mathbb{R}$, $\xi \to 1^{-}$).
Hence  if we show that the determinant $D \leq 0$, then from the positivity of $P^{\dagger}(z|z;\xi)$ it follows 
that $\Delta   \geq 0$. From (\ref{eq:043}) in the $\xi \to 1^{-}$ limit:
$-D= 1 + w - (u\alpha + v\beta)$, where we used the normalization $f+g=1$.
Let us examine the sign of this expression. Using the definitions from (\ref{eq:044}), we obtain:
\begin{equation}
-D =1 +a \alpha (1-g) +d \beta (1-f) -\alpha fb -\beta gc. \label{prove3}
\end{equation}
Recall, that $f+g =1$, with $f$ and $g$ being probabilities chosen arbitrarily. 
Then $D$ can be rewritten as
\begin{equation}
-D = 1+ f\alpha(a-b) + g \beta (d-c). \label{prove4}
\end{equation}
Determinant $D$ also appears in the expressions of the site occupancy generating functions $P^{\dagger}(s|s;\xi)$ and $P^{\dagger}(s|s'; \xi)$, in (\ref{eq:049}).  In the $\xi \to 1^{-}$ limit, these expressions take the form of
\begin{eqnarray}
&& P^{\dagger}(s|s; 1^{-}) = -\frac{1}{D} (b+ g \beta t), \label{eq:elso} \\
&& P^{\dagger}(s|s'; 1^{-}) = -\frac{1}{D} (a+ g \beta t). \label{eq:mas}
\end{eqnarray}
The sum $cP^{\dagger}(s|s; 1^{-}) + d P^{\dagger}(s|s';1^{-})$ is always non-negative 
($c,d \geq 0$), implying  
\begin{equation}
cP^{\dagger}(s|s;1^{-}) + d P^{\dagger}(s|s';1^{-}) = 
-\frac{1}{D} \left[ ad+ bc + t \beta g (c+d)\right] \geq 0. \label{prove5}
\end{equation} 
As $a, b, c, d \geq 0$ (for $\xi \to 1^{-}$, $\xi \in \mathbb{R}$), $0 \leq \beta \leq 1$, and $g \in [0,1]$  is an arbitrarily chosen transition probability, for any given $\xi$, $g$ can always be chosen 
to be small enough, such that $ad+ bc + t\beta g (c+d) \geq 0$, independently on the sign of $t$.
(Note that  $a, b, c, d$ are independent on $g$ or $f$).
Thus, for small $g$ values, $D \leq 0$, and based on (\ref{pidagdelta}) this implies that $\Delta \geq 0$.
Since $\Delta$ is {\em independent of} $f$ and $g$, this also implies that for arbitrary $f$ ($g=1-f$),
we have:
\begin{eqnarray}
&&-D =1 + f\alpha (a-b) + g\beta (d-c) \geq 0,  \;\; \text{as}\;\;\; \xi \to 1^{-}\label{eq:prove6} \\
&&\Delta = 1+ (a\alpha + d\beta) + (ad-bc) \alpha \beta \geq 0, \;\; 
\text{as}\;\;\; \xi \to 1^{-}\label{eq:prove7} 
\end{eqnarray}
Hence, if we choose $f = 1$, $g = 0$ in (\ref{eq:prove6}), followed by $f = 0$, $g = 1$ we obtain:
\begin{eqnarray}
&&1+ \alpha(a-b) \geq 0, \label{prove8a} \\
&&1+ \beta (d-c) \geq 0, \label{prove8b}
\end{eqnarray}
as $\xi \to 1^{-}$. Observe that (\ref{prove8b}) is also obtained by
switching $s$, $s'$, to $s'$ and $s$ respectively, in (\ref{prove8a}) (which holds for any $s$ and $s'$).
Using (\ref{eq:005}), and the fact that $F(s|s_0;\xi) \leq 1$ even at 
$|\xi| = 1$ ($F(s|s_0;1) = R(s|s_0)$ is the {\em probability} that $s$ is ever reached from $s_0$), after replacing $s_0$ by $s' \neq s$, we find that in the limit $\xi \to 1^{-}$
\begin{equation}
 a-b \leq 0\;. \label{basic1}
\end{equation}
However, although $a \leq b$, the difference $a-b$ cannot be arbitrarily negative, as shown in
(\ref{prove8a}): $a-b \geq - \alpha^{-1}$. Similarly, it holds:
\begin{equation}
 d-c \leq 0\;.  \label{basic2}
\end{equation}
An immediate consequence of these inequalities, is that in the case of recurrent walks, where 
both $a$ and $b$ (and similarly, $c$ and $d$) diverge as $\xi \to 1^{-}$, their difference is 
nevertheless bounded. From (\ref{prove8b}) and (\ref{eq:prove7}) it follows that every term in
the expression of $\overline{W}(s;1^{-})$ is non-negative, and thus $\overline{W}(s;1^{-}) \geq 0$.

\medskip
Next we show that $\overline{W}(s;1^{-}) < \infty$ (finite). Recall, that $1 \leq X_n \leq n$, that is, the number of edges cannot grow faster than linearly with 
time (at most one new edge can be discovered in a time step). This inequality written in terms of generating functions becomes:
\beq{prove12}
\frac{1}{1-\xi} \leq  X(\xi) \leq \frac{\xi}{(1-\xi)^2}\;,\;\;\;\xi \in \mathbb{R},\;\; \xi > 0\;.
\eeq
Let us assume that on the contrary, 
$\overline{W}(s;\xi) \to \infty$ as $\xi \to 1^{-}$. This means that for 
{\em any arbitrarily large constant} $C$, there is a $\xi_0 \in \mathbb{R}$, $\xi_0 < 1$, such that  $\overline{W}(s;\xi) > C$ for all $\xi > \xi_0$, $\xi \to 1^{-}$. Thus:
\begin{equation}
X(\xi)=\dfrac{\xi}{1-\xi} \sum_{s} \overline{W}(s;\xi) P(s|s_0;\xi) > \dfrac{C \xi}{1-\xi} 
 \sum_{s} P(s|s_0;\xi) = \dfrac{C \xi}{(1-\xi)^2}, \label{contra1}
 \end{equation}
where in the last step we used the identity $\sum_s P(s|s_0) = (1-\xi)^{-1}$, which is a direct 
consequence of the normalization condition $\sum_s P_n(s|s_0) = 1$). Since $C$ is arbitrarily large, it can certainly be chosen such that $C > 1$, and thus (\ref{contra1}) will be contradicting (\ref{eq:prove12}).

Let us write (\ref{eq:057}) in the form:
\beq{xxww}
\overline{W}(s;\xi) =\sum_{s'}\alpha \frac{1+(d-c)\beta \xi}{1+(a \alpha +d \beta)\xi + (ad-bc) \alpha \beta \xi^{2}} =\sum_{s'} \alpha \; \theta(s,s',\xi) \;.
\eeq
Functions $a$ and $d$ can be expressed in terms of the first-passage time generating function by using relation (\ref{eq:005}): $a=F(s|s';\xi) P(s|s;\xi)=\eta b$, $d=F(s'|s;\xi) P(s'|s';\xi) = \eta'c$, where we have introduced the notations $\eta= F(s|s';\xi) \leq 1$ and $\eta' = F(s'|s;\xi) \leq 1$. In the $\xi \to 1^-$ limit, we obtain for $\theta(s,s',1^-)$
\beq{theta1}
\theta(s,s',1^-) =\frac{1- (1-\eta') c \beta}{ 1+ b\alpha \eta + c \beta \eta'- bc (1- \eta \eta')\alpha \beta}\;.
\eeq
The denominator can be written as: $1+ b\alpha \eta + c \beta \eta'- bc (1- \eta \eta')\alpha 
\beta = [ 1-(1-\eta')c \beta] [1+(1+\eta)b \alpha] - b \alpha (1+c\beta \eta') +c \beta (1+b\alpha \eta)$. 
Let  $u'=1 - (1-\eta')c\beta$ and $u = 1-(1-\eta)b\alpha$. Since $\eta'c = d$ and $\eta b = a$, we have
$u'=1+\beta(d-c)$ and $u = 1+\alpha (a-b)$. Thus, from (\ref{prove8a}-\ref{prove8b}) it follows that
$u' \geq 0$ and $u \geq 0$. Moreover, we can write $1-\eta' = (1-u')/(c\beta)$, $1-\eta = (1-u)/(b\alpha)$. 
Since $1-\eta' \geq 0$ and $1-\eta \geq 0$ (see above) and as both $c\beta$ and $b\alpha$ are positive
it follows that one must have:
\begin{equation}
0 \leq u' \leq 1\;,\;\;\;\;0 \leq u \leq 1\;.  \label{ins}
\end{equation}
Using the $u'$ notation, equation \refeq{theta1} is written as:
\beq{theta2}
\theta(s,s',1^-) =\frac{u'}{u' [1+(1+\eta)b \alpha] +c \beta (1+b\alpha \eta) -b \alpha (1+c\beta \eta ')}\;.
\eeq
In the denominator we then use $1+ c\beta \eta' = u'+ c\beta$. After the cancelations we find:
\beq{theta3}
\theta(s,s',1^-) = \frac{u'}{u'(1+b\alpha \eta) +c\beta(1+b \alpha \eta) -bc\alpha \beta} =
\frac{u'}{uu' + u'b \alpha + u c \beta}\;.
\eeq
The last equality was obtained after using $1+ b\alpha \eta = u + b\alpha$.
Hence: 
\begin{equation}
\overline{W}(s;1^-) = \frac{1}{b}\sum_{s'} \alpha \; \frac{u'}{u'\alpha + u \varphi \beta + u u' / b},
\end{equation}
where $\varphi = c/b = P(s'|s';1^-)/P(s|s;1^-)$. 
Taking into account (\ref{ins}), the terms in the sum are all positive and finite (even if $b \to \infty$), 
that is there is 
$C > 0$ constant such that:
\beq{theta4}
\overline{W}(s;1^-) <  \frac{1}{b}\sum_{s'} \alpha \; C = \frac{1}{b}\;C.
\eeq
(Recall, that due to normalization $\sum_{s'} \alpha = \sum_{s'} p(s'|s) = 1$.)
This implies from (\ref{eq:056}) that in the limit $\xi \to 1^-$ ($\xi$ is very close to $1$ but not quite $1$):
\begin{equation}
X(s_0;\xi) < \frac{1}{1-\xi} \sum_{s}C \frac{P(s|s_0;\xi)}{b} = C S(s_0;\xi)
\end{equation}
Clearly, after the first loop was made by the walker (which happens after long enough times,
otherwise we have $X_n = S_n - 1$ identically, in which case trivially $\mu = \lambda$), we have 
$S_n < X_n$ and hence:
\begin{equation}
S(s_0;\xi) < X(s_0;\xi) < C S(s_0;\xi)\;,\;\;\;\mbox{as}\;\;\;\xi \to 1^-\;,
\end{equation}
showing that the leading order behavior for $\langle X_n \rangle$ and $\langle S_n \rangle$ 
are identical, i.e., $\mu = \lambda$.

\medskip

\begin{figure}[htb]
  \begin{center}
    \centerline{\includegraphics[scale=0.55]{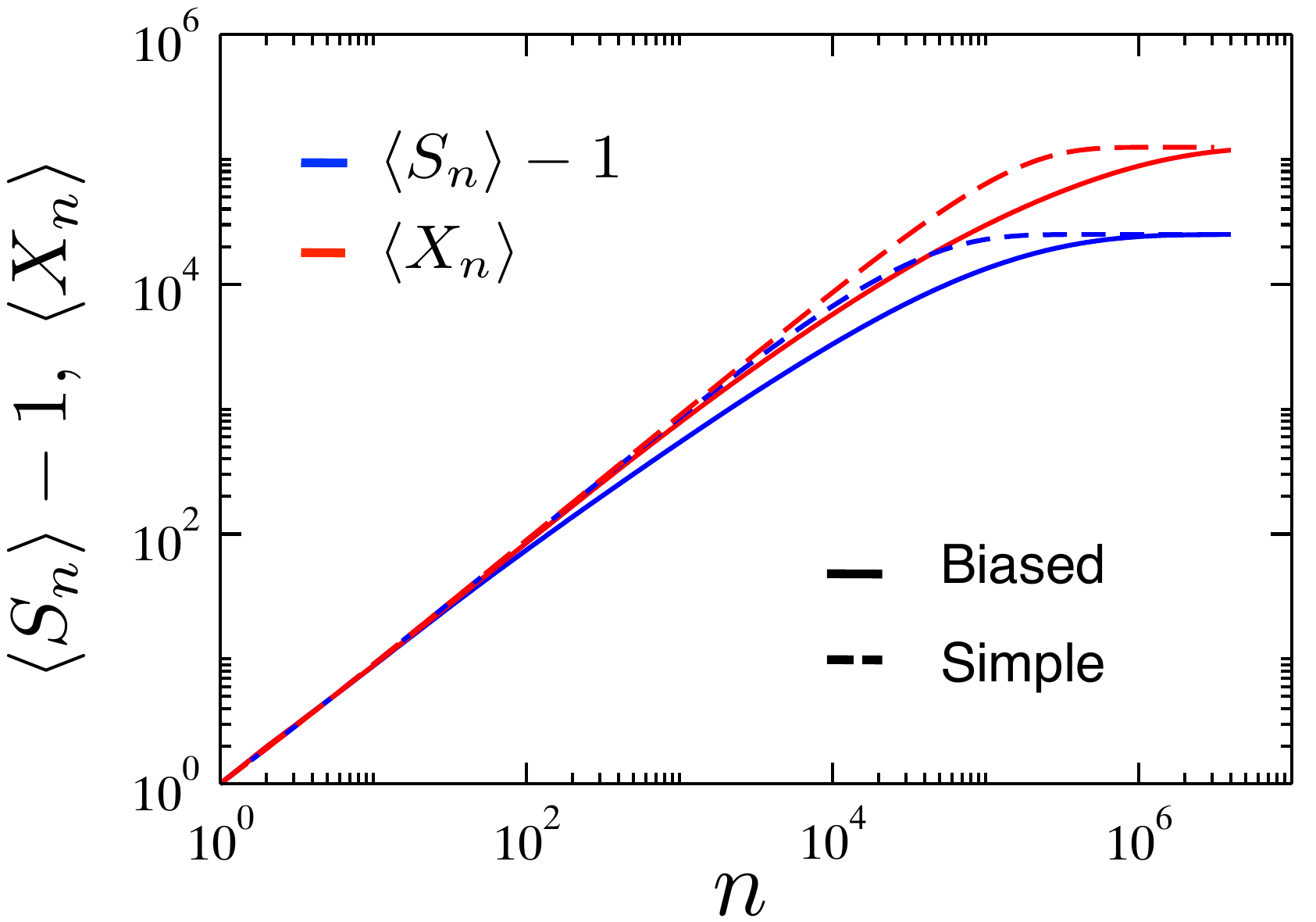}}
    \caption{Comparison of the average number of discovered nodes (blue lines) and edges (red lines) on scale-free BA model ($N=25000$, $\av{k}=10$), for simple random walk (dashed lines) and for a walk that is biased towards high degree nodes (solid lines). }
    \label{fig:biased_walk_scalefree}
  \end{center}
\end{figure}

\noindent {\bf Discussion.}
Recall, that $\av{S_n} \sim n^{\lambda}$ ($\av{X_n} \sim n^{\mu}$) means  $\av{S_n} \simeq n^{\lambda} L(n)$ ($\av{X_n} \simeq n^{\mu}F(n)$) as $n \to \infty$, where $L(n), F(n)$ are slowly varying functions, that is  $L(\zeta x)/L(x) \to 1$ ($F(\zeta' x)/F(x) \to 1$) when $x \to \infty$, for any $\zeta, \zeta' >0$. 
\begin{enumerate}
\item[i)] \singlespace When $\mu=\lambda \neq 0$, $\av{S_n}$ and $\av{X_n}$ obey the same scaling laws, however for early times, the corrections $L(n)$ and $F(n)$ can be  dominant. This results in a slight deviation between the two curves, e.g., Sierpinski gasket, square lattice, as seen in the main article. For large graphs, in the $n \to \infty$ limit, these corrections diminish.
\item[ii)] When $\mu=\lambda=0$, the growth of $\av{S_n}$ is characterized by the function $L(n)$, while the growth of $\av{X_n}$ is dictated by the $F(n)$, which are not necessarily the same. This case is shown in \reffig{biased_walk_scalefree} for the scale-free BA model. For simple random walks, $\av{S_n}$ and $\av{X_n}$ grow at the same rate, while for a walk that is biased towards high degree nodes, after a short linear growth region they slowly grow following different curves 
till they saturate.  For the biased STP walk towards 
higher degrees we took $p(s'|s) = k_s \left(\sum_{s''\in \langle s \rangle} k_{s''} \right)^{-1}$ where
$k_s$ is the degree of node $s$ and  $\langle s \rangle $ denotes the set of graph neighbors of $s$.
\end{enumerate}


\begin{thebibliography}{10}

\bibitem{RandWalksRandEnvVol1_H95}
Barry~D. Hughes.
\newblock {\em Random Walks and Random Environments}, volume 1: Random Walks.
\newblock Oxford U. P., 1995.

\bibitem{IntrProbTheorApplVol2_F71}
William Feller.
\newblock {\em An Introduction to Probability Theory and its Applications},
  volume~2.
\newblock Wiley, 2 edition, 1971.

\bibitem{RevMarkChRandWalksGraphs_AF99}
David~J. Aldous and James~Allen Fill.
\newblock {\em Reversible Markov Chains and Random Walks on Graphs}.
\newblock 1999.

\bibitem{CombCompGeom_V05}
Santosh Vempala.
\newblock Geometric random walks: A survey.
\newblock {\em Comb. Comp. Geom.}, 52:573--612, 2005.

\bibitem{Proc8IntWWW_HHMN99}
Monika~R. Henzinger, Allan Heydon, Michael Mitzenmacher, and Mark Najork.
\newblock In {\em Proc. of the 8th Intern. World Wide Web Conf., Toronto,
  Canada}, page 213. Elsevier Science, May 1999.

\bibitem{Holme_traffic03}
Petter Holme.
\newblock Congestion and centrality in traffic flow on complex networks.
\newblock {\em Adv. Comp. Sys.}, 6:163--176, 2003.

\bibitem{PRL_Rieger04}
Jae~Dong Noh and Heiko Rieger.
\newblock Random walks on complex networks.
\newblock {\em Phys. Rev. Lett.}, 92(11):118701, 2004.

\bibitem{PRL_epidemic01}
Romualdo Pastor-Satorras and Alessandro Vespignani.
\newblock Epidemic spreading in scale-free networks.
\newblock {\em Phys. Rev. Lett.}, 86(14):3200--3203, 2001.

\bibitem{Phys_Rev_E_N02}
Mark~E.J. Newman.
\newblock Spread of epidemic disease on networks.
\newblock {\em Phys. Rev. E}, 66:016128, 2002.

\bibitem{PRE_Volchenkov02}
D.~Volchenkov, L.~Volchenkova, and Ph. Blanchard.
\newblock Epidemic spreading in a variety of scale free networks.
\newblock {\em Phys. Rev. E}, 66:046137, 2002.

\bibitem{PRE_Adamic01}
Lada~A. Adamic, Rajan~M. Lukose, Amit~R. Puniyani, and Bernardo~A. Huberman.
\newblock Search in power-law networks.
\newblock {\em Phys. Rev. E}, 64:046135, 2001.

\bibitem{Strogatz_Nature01}
Steven~H. Strogatz.
\newblock Exploring complex networks.
\newblock {\em Nature}, 410:268, 2001.

\bibitem{StatMech_BA02}
R\'eka Albert and Albert-L\'aszl\'o Barab\'asi.
\newblock Statistical mechanics of complex networks.
\newblock {\em Rev. Mod. Phys.}, 74:47--97, 2002.

\bibitem{DvoEr51}
A.~Dvoretzky and P.~Erd\H{o}s.
\newblock In J.~Neyman, editor, {\em Proc. of the 2nd Berkeley Symp. on Math.
  Stat. Prob.}, pages 353--367. U. California Press, Berkeley, 1951.

\bibitem{MW65}
Elliott~W. Montroll and George~H. Weiss.
\newblock Random walks on lattices. {I}{I}.
\newblock {\em J. Math. Phys.}, 6:167--181, 1965.

\bibitem{Cassi_review05}
R.~Burioni and D.~Cassi.
\newblock Random walks on graphs: idea, techniques and results.
\newblock {\em J. Phys. A: Math. Gen.}, 38:R45--R78, 2005.

\bibitem{Lovasz_93Survey}
L.~Lov\'{a}sz.
\newblock Random walks on graphs: a survey.
\newblock {\em Combinatorics, Paul Erd\H{o}s is Eighty}, 2:1--46, 1993.

\bibitem{PRE_Lahtinen01}
Jani Lahtinen, J\'{a}nos Kert\'{e}sz, and Kimmo Kaski.
\newblock Scaling of random spreading in small world networks.
\newblock {\em Phys. Rev. E}, 64:057105, 2001.

\bibitem{PRE_Eivind03}
E.~Almaas, R.~V. Kulkarni, and D.~Stroud.
\newblock Scaling properties of random walks on small-world networks.
\newblock {\em Phys. Rev. E}, 68:056105, 2003.

\bibitem{EdgeEx_CS_Feige93}
Greg Barnes and Uriel Feige.
\newblock Short random walks on graphs.
\newblock {\em Annual ACM Symposium on Theory of Computing}, pages 728--737,
  1993.

\bibitem{EPL_Ram07}
A.~Ramezanpour.
\newblock Intermittent exploration on a scale-free network.
\newblock {\em Europhys. Lett.}, 77:60004, 2007.

\bibitem{Chung1960}
K.~L. Chung.
\newblock {\em Markov Chains with Stationary Transition Probabilities}.
\newblock Springer-Verlag, Berlin, 1960.

\bibitem{ErdosRgraph}
P.~Erd\H{o}s and A.~R\'{e}nyi.
\newblock On random graphs {I}.
\newblock {\em Publ. Math.}, 6:290--297, 1959.

\bibitem{PREDall02_rgg}
Jesper Dall and Michael Christensen.
\newblock Random geometric graphs.
\newblock {\em Phys. Rev. E}, 66:016121, 2002.

\bibitem{Science_BA99}
Albert-L\'aszl\'o Barab\'asi and R\'eka Albert.
\newblock Emergence of scaling in random networks.
\newblock {\em Science}, 286:509--512, October 1999.

\bibitem{Ravasz_Hier02}
Erzs\'ebet Ravasz, A.~L. Somera, D.~A. Mongru, Zolt\'an~N. Oltvai, and
  Albert-L\'aszl\'o Barab\'asi.
\newblock Hierarchical organization of modularity in metabolic networks.
\newblock {\em Science}, 297:1551, 2002.

\bibitem{PNASGallos07}
Lazaros~K. Gallos, Chaoming Song, Shlomo Havlin, and Hern\'{a}n~A. Makse.
\newblock Scaling theory of transport in complex biological networks.
\newblock {\em PNAS}, 104(19):7746--7751, May 2007.

\bibitem{Auriac_JPA_83}
J.~C.~Angles d'Auriac, A.~Benoit, and R.~Rammal.
\newblock Random walk on fractals: numerical studies in two dimensions.
\newblock {\em Jour. Phys. A}, 16:4039--4051, 1983.

\bibitem{Koenig01ants}
Sven Koenig, Boleslaw Szymanski, and Yaxin Liu.
\newblock Efficient and inefficient ant coverage methods.
\newblock {\em Annals Math. \& Art. Int.}, 31:41--76, 2001.

\bibitem{WilfGfunction05}
H.~S. Wilf.
\newblock {\em generatingfunctionology}.
\newblock A. K. Peters LTD, 3rd edition, 2005.

\end{thebibliography}

\end{document}